\documentclass[pdflatex,sn-mathphys-num]{sn-jnl}

\usepackage{graphicx}%
\usepackage{multirow}%
\usepackage{amsmath,amssymb,amsfonts}%
\usepackage{amsthm}%
\usepackage{mathrsfs}%
\usepackage[title]{appendix}%
\usepackage{xcolor}%
\usepackage{textcomp}%
\usepackage{manyfoot}%
\usepackage{booktabs}%
\usepackage{algorithm}%
\usepackage{algorithmicx}%
\usepackage{algpseudocode}%
\usepackage{listings}%

\usepackage{hyperref}

\usepackage{csquotes}
\usepackage{xcolor}
\usepackage[colorinlistoftodos,prependcaption]{todonotes}
\usepackage[inline]{enumitem}
\usepackage{xspace}
\usepackage{subcaption}
\usepackage{longtable}
\usepackage{afterpage}
\usepackage{amssymb}
\usepackage[nohyperlinks,printonlyused]{acronym}
\usepackage{amsmath}
\usepackage{pifont}
\newcommand{\cmark}{\ding{51}}%
\newcommand{\xmark}{\text{\ding{55}}}
\usepackage[T1]{fontenc}

\usepackage{float}
 \usepackage[usenames,dvipsnames]{pstricks}
\usepackage{pstricks-add}
 \usepackage{pst-grad} 
 \usepackage{pst-plot} 
 \usepackage[space]{grffile} 
 \usepackage{etoolbox} 
 \makeatletter 
 \patchcmd\Gread@eps{\@inputcheck#1 }{\@inputcheck"#1"\relax}{}{}
 \makeatother
\usepackage{tikz}
\usetikzlibrary{mindmap,trees}
\usepackage{verbatim}
\usepackage{xspace}
\usepackage{multirow} 
\usepackage{booktabs}
\usepackage{multicol}
\usepackage{float}
\usepackage{array}
\usepackage{amssymb}
\usepackage{pifont}

\usepackage{algpseudocode}
\usepackage{mathtools}
\usepackage{float}

\usepackage{geometry}
\theoremstyle{thmstyleone}%
\theoremstyle{thmstyletwo}%

\theoremstyle{thmstylethree}%

\begin{document}

\title[Deep Transfer Learning Techniques for Kidney Cancer Diagnosis]{Deep Transfer Learning Techniques for Kidney Cancer Diagnosis}


\author[1]{\fnm{Yassine} \sur{Habchi}}\email{habchi@cuniv-naama.dz}


\author[2]{\fnm{Hamza} \sur{Kheddar}}\email{kheddar.hamza@univ-medea.dz}


\author[3]{\fnm{Yassine} \sur{Himeur}}\email{yhimeur@ud.ac.ae}


\author*[4]{\fnm{Mohamed Chahine } \sur{Ghanem}}\email{mohamed.chahine.ghanem@liverpool.ac.uk}

\author[5]{\fnm{Abdelkrim} \sur{Boukabou}}\email{aboukabou@univ-jijel.dz}


\author[6]{\fnm{Shadi} \sur{Atalla}}\email{satalla@ud.ac.ae}


\author[7]{\fnm{Wathiq} \sur{Mansoor}}\email{wmansoor@ud.ac.ae}


\author[8]{\fnm{Hussain} \sur{Al-Ahmad}}\email{halahmad@ud.ac.ae}


\affil[1]{\orgdiv{Institute of Technology}, \orgname{University Center Salhi Ahmed}, \orgaddress{\postcode{45000}, \state{Naama}, \country{Algeria}}}

\affil[2]{\orgdiv{LSEA Laboratory, Electrical Engineering Department, Faculty of Technology}, \orgname{University of Medea}, \orgaddress{ \postcode{26000}, \state{Medea}, \country{Algeria}}}

\affil[3, 6, 7, 8]{\orgdiv{College of Engineering and Information Technology}, \orgname{University of Dubai}, \orgaddress{\street{Dubai}, \country{UAE}}}

\affil*[4]{\orgdiv{Department of Computer Science}, \orgname{University of Liverpool}, \orgaddress{\postcode{L69 3BX}, \city{Liverpool}, \country{UK}}}

\affil[5]{\orgdiv{Department of Electronics}, \orgname{University of Jijel}, \orgaddress{\city{Jijel}, \postcode{18000}, \country{Algeria}}}


\abstract{Incurable diseases continue to pose significant challenges to global healthcare systems, with their prevalence being influenced by lifestyle choices, economic conditions, social factors, and genetic predispositions. Among these, kidney disease remains a critical health issue, affecting individuals worldwide and necessitating continuous research to enhance early diagnosis and treatment strategies. In recent years, deep learning (DL) has gained prominence in medical imaging and diagnostics, offering significant advancements in automatic kidney cancer (KC) detection. However, the effectiveness of DL models largely depends on the availability of high-quality medical datasets, which are often scarce, expensive, and domain-specific. Additionally, these models require substantial computational resources and storage capacity, limiting their applicability in real-world clinical settings.  To address these challenges, transfer learning (TL) has emerged as a promising solution, enabling the adaptation of pre-trained models from different domains to improve KC diagnosis. This paper presents the first comprehensive survey of DL-based TL frameworks specifically applied to KC detection. By systematically reviewing existing approaches, this study identifies key methodologies, explores their advantages and limitations, and critically analyzes their effectiveness in clinical applications. Furthermore, the paper highlights current challenges in implementing TL for medical imaging and discusses emerging trends that could shape future research in this field. Ultimately, this review underscores the transformative role of TL in precision medicine, particularly in oncology, by enhancing diagnostic accuracy, reducing computational costs, and facilitating the deployment of AI-driven healthcare solutions. The findings of this study provide valuable insights for researchers and practitioners, paving the way for further advancements in KC diagnostics and personalized treatment approaches.}

\keywords{Medical image, Kidney cancer diagnosis, Deep learning, Transfer learning, Domain adaptation, Fine-tuning.}



\maketitle

\section{Introduction}  \label{intro}

\begin{figure*}[t!]
\centering
\includegraphics[scale=0.6]{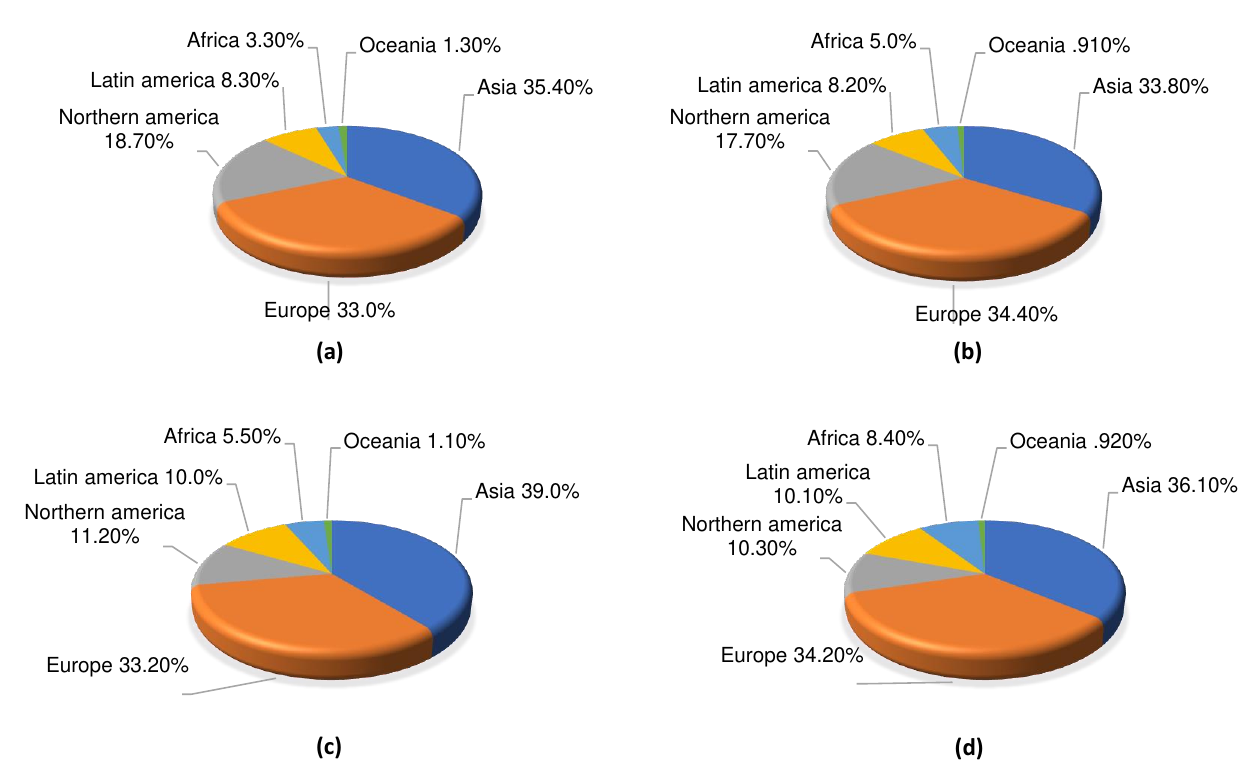}
\caption{{The estimated number of KC incidence and mortality in the world.}}
\label{fig1}
\end{figure*}

The distinction between tumor subtypes and pathological grades is a critical aspect of cancer diagnosis and treatment planning. Tumor subtypes classify the tumor based on its cellular and molecular characteristics. For instance, in \ac{RCC}, the major subtypes—\ac{CCRCC}, \ac{PRCC}, and \ac{CRCC}—differ in their cellular appearance, growth patterns, and genetic alterations. These subtypes have distinct biological behaviors, prognoses, and responses to treatment. \ac{RCC} is a cancer that impacts kidney function, which plays a vital role in regulating the body’s fluid balance by filtering out toxins and excreting waste products. Disorders such as \ac{KC} can impair this function \citep{lommen2021diagnostic}. As shown in Figure \ref{fig1}, the estimated incidence and mortality rates for \ac{KC} in 2022, for both genders globally. In terms of pathological grading, \ac{KC} is classified on a scale of 1 to 4, with 1 being the least aggressive and 4 being the most aggressive. This grading is determined by analyzing cancer cells under a microscope to assess the degree of differentiation and anaplasia. Grade 1 (\ac{PRCC}) tumors are well-differentiated, closely resembling normal kidney cells and grow slowly \citep{semko2024comparison}. Grade 2 (\ac{CRCC}) tumors are moderately differentiated with some resemblance to normal kidney cells, growing more slowly than grades 3 and 4. Grade 3 tumors (\ac{CCRCC}) consist of poorly differentiated cells that bear little resemblance to normal kidney cells and tend to grow and spread rapidly \citep{zheng2021deep}. Grade 4 tumors (\ac{SRCC}) are highly aggressive and rare, growing and spreading rapidly. Pathological grading, which reflects how abnormal the tumor cells are compared to normal tissue, helps predict the tumor’s aggressiveness and guides therapeutic decisions. While subtypes categorize the tumor based on its origin and intrinsic characteristics, grading evaluates the degree of malignancy. Understanding both the subtypes and grades is crucial for determining prognosis and selecting the most effective treatment strategy. We have also referenced the WHO pathological classification and grading system for renal tumors \citep{goswami20242022}, providing a reliable and standardized framework for understanding these concepts. These revisions enhance the scientific rigor and clinical relevance of our paper, making it more valuable for researchers and healthcare professionals alike.

\begin{table*}[ht!]
\centering
\section*{\small{Acronyms and Abbreviations}}
\begin{multicols}{3}
\footnotesize
\begin{acronym}[AWGN]  
\acro{AI}{artificial intelligence} 
\acro{BERT}{bidirectional encoder representations from Transformer}
\acro{CNN}{convolutional neural network}
\acro{CT}{computerized tomography}
\acro{CTC}{connectionist temporal classification}
\acro{CRCC}{chromophobe renal cell carcinoma} 
\acro{CCRCC}{clear cell renal cell carcinoma} 
\acro{CAD}{computer-aided detection}
\acro{DA}{domain adaptation}
\acro{DG}{domain generalization}
\acro{DL}{deep learning}
\acro{FSL}{few-shot learning}
\acro{FL}{federated learning}
\acro{KC}{kidney cancer}
\acro{KiTS19}{kidney tumor segmentation 19}
\acro{LRP}{layer-wise relevance propagation}
\acro{LLM}{large language model}
\acro{ML}{machine learning}
\acro{MRI}{magnetic resonance imaging}
\acro{PRCC}{papillary renal cell carcinoma}
\acro{RCC}{renal cell carcinomas}
\acro{RDAN}{reciprocal domain adaptation network}
\acro{SVM}{support vector machine}
\acro{SRCC}{sarcomatoid renal cell carcinoma}
\acro{SSL}{self-supervised learning}
\acro{TL}{transfer learning}
\acro{TCGA}{the cancer genome atlas}
\acro{US}{ultrasound}
\acro{UL}{unsupervised learning}
\acro{XAI}{explainable AI}
\end{acronym}

\end{multicols}
\end{table*}

Moreover, KC can be identified by employing a range of imaging methods such as \ac{MRI}, X-rays, and \ac{US} imaging \citep{pimentel2021use}. These methods are capable of generating comprehensive visuals of the kidneys, enabling physicians to spot any irregularities or tumors \citep{caroli2021basic,habchi2025advanced}. Nonetheless, despite the utility of medical imagery in disease diagnosis, its reliability is not absolute in certain scenarios. Particularly in cancer cases, more precise techniques may be required to aid physicians in detecting malignant cells and assessing their metastasis to other body regions \citep{zhou2020imaging}. Moreover, interpreting whether an image truly signifies the presence of cancer can be challenging. Medical imagery is often rich in detail and data, complicating its analysis. In addition, the overwhelming volume of images and visual assessments may contribute to medical inaccuracies. The burden of analyzing numerous images can diminish a radiologist's focus and their capacity to notice crucial nuances. Such circumstances can lead to misdiagnosis, diagnosis delays, or even inappropriate treatments \citep{zhou2020imaging}.

In recent years, \ac{DL}-based early detection methods have recorded significant achievements in identifying various terminal illnesses. These methods focus on accurately extracting diverse features that assist medical professionals in the diagnostic process through the utilization of the network's multiple layers \citep{zhang2023deep,hamza2023hybrid}. Attaining the necessary diagnostic precision for incurable diseases affecting the kidneys presents challenges due to: (i) the insufficiency of data resources for the training phase \citep{himeur2022deep,wu2023application}; (ii) the time-consuming, expensive, and exhaustive nature of data collection \citep{arif2023enhancing,subramanian2023multiple}; (iii) the requirement for data annotation in the training phase; and (iv) the impracticality of manually labeling large datasets by experts, which is often unfeasible \citep{rasmussen2022artificial,bechar2023harnessing}.

While many \ac{DL} algorithms, such as the \ac{CNN} \citep{liang2023srenet,habchi2023ai1,kheddar2022high, djeffal2023noise}, are proficient at recognizing complex patterns, they encounter significant challenges, particularly during the transition from training to testing phases, which can impede their effectiveness. Additionally, various studies \citep{gharaibeh2022radiology} have pointed out that the majority of datasets utilized for diagnosing either early or advanced cancer stages are often inadequate for the diagnostic process. This inadequacy is due to the intricate structure of certain organs, like the kidneys, and the variations in image depth, making it exceedingly difficult to accurately identify the type or grade of cancer \citep{hadjiyski2020kidney,da2020kidney,wu2020automated}. Furthermore, these algorithms heavily rely on the assumption that the training and testing data are extracted from identical feature spaces and distributions. A deviation between these can necessitate a comprehensive model reconfiguration, which is not only costly but also labor-intensive. As \citep{hussain2020volumetric} notes, supervised learning algorithms can yield remarkable outcomes with extensive volumes of annotated data. Yet, they generally falter with the complex variations present in medical images, where labeled data are scarce, and backgrounds are complex. Conversely, diagnostic systems powered by \ac{AI} critically depend on having access to well-distributed and annotated datasets for training, a requirement that is often hard to fulfill \citep{chen2022novel}. This underscores the need for enhancements in \ac{AI}-based diagnostic systems for incurable diseases, aiming to simplify them to improve diagnostic efficiency, timing, and precision \citep{gottlich2023effect}. 

\begin{figure*}[t]
\begin{center}
\includegraphics[width=0.9\columnwidth]{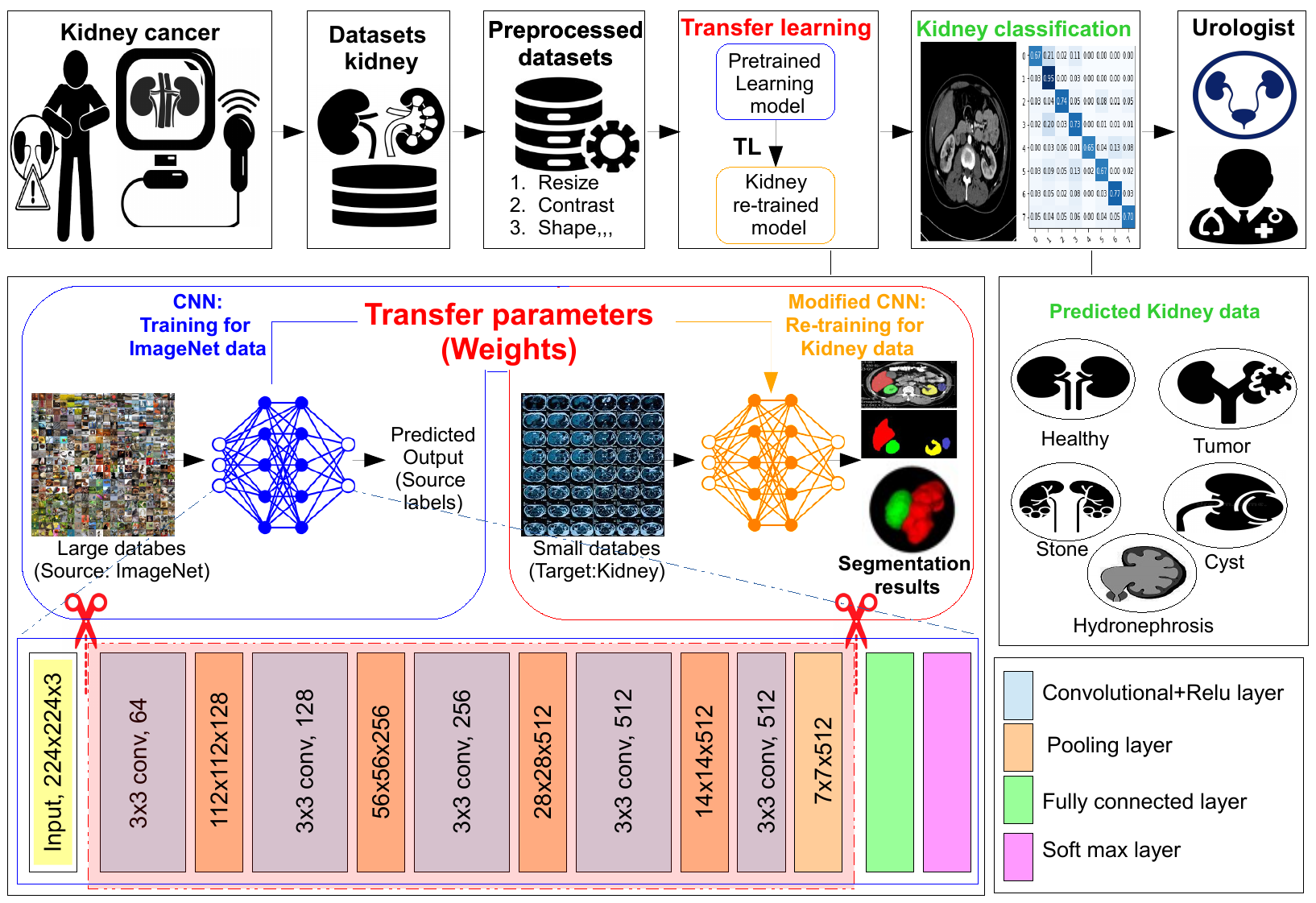}
\end{center}
\caption{Principle of Kidney classification based on TL.}
\label{fig2}
\end{figure*}

To meet these goals, efforts are being directed towards leveraging \ac{TL} in many research fields \citep{kheddar2023deepASR,himeur2023video,kheddar2023deep,mazari2023deep,kheddar2024automatic}. This process fundamentally involves repurposing a network that has been previously trained on a specific dataset for a distinct task. This entails employing established models such as LeNet, AlexNet, VGGNet, ResNet, GoogLeNet, DenseNet, XceptionNet, and SqueezeNet \citep{himeur2023video} for tasks on smaller datasets, thereby facilitating the omission of labels. This concept draws inspiration from the human ability to apply knowledge and experiences from past learning to foster new insights. Consequently, various databases hosting extensive image collections have been made available online, such as the \ac{CNN} algorithm's application on the ImageNet dataset, which is accessible for public experimentation with \ac{DL}-based algorithms. This approach effectively mitigates data scarcity issues and circumvents the need for extensive data annotation \citep{himeur2023face}. Moreover, this algorithmic approach alleviates the lengthy durations typically required for learning processes, which can extend over days or weeks. In the realm of diagnosing terminal illnesses, \ac{TL} has been categorized into two methodologies: fine-tuning and feature extraction, both of which have shown promising outcomes, notably in \ac{KC} detection. Numerous studies within this domain, for instance, \citep{nasir2022kidney}, exemplify its application. \ac{TL} is particularly beneficial for cancer detection and diagnosis, including \ac{KC} \citep{al2023automated, elharrouss2021panoptic}. Utilizing a pre-trained DL model, like a \ac{CNN}, fine-tuned on kidney imaging scans (e.g., \ac{CT} or \ac{MRI}), can enhance its efficacy in identifying and diagnosing kidney tumors (Figure \,\ref{fig2}). A method for fine-tuning a model for \ac{KC} detection might involve using a dataset comprising both healthy and cancerous kidney scans, adjusting the model's final layers' weights to better recognize cancerous areas. Additionally, the model could be optimized by either freezing specific layers while training the last few or starting a new one with the pre-trained weights. An alternative \ac{TL} strategy in \ac{KC} detection could leverage a model pre-tuned for a similar task, such as lung cancer detection, and further fine-tune it on kidney scans, beneficial when large datasets for \ac{KC} detection are unavailable. \ac{TL} proves invaluable in medical imaging, significantly when large, labeled datasets are scarce, enhancing the efficiency and accuracy of kidney tumor identification and diagnosis.

{This study is the first comprehensive review based on \ac{DL} and \ac{TL} for the diagnosis of \ac{KC}. While many studies have explored the application of \ac{DL} in medical imaging, few have systematically addressed the integration of \ac{TL} specifically for \ac{KC} detection. Previous reviews either focus on \ac{DL} alone or do not consider \ac{TL} frameworks, which are particularly important for overcoming challenges like dataset scarcity and limited computational resources in real-world clinical settings. This paper fills that gap by presenting a thorough survey of \ac{DL}-based \ac{TL} frameworks applied to \ac{KC} detection, systematically analyzing existing approaches and identifying their advantages and limitations. It goes further by addressing challenges such as the availability of high-quality medical datasets, the need for extensive computational resources, and the difficulties in adapting \ac{DL} models for practical use in clinical environments. Additionally, the paper highlights emerging trends in \ac{TL}, which could play a crucial role in enhancing diagnostic accuracy and facilitating \ac{AI}-driven healthcare solutions. Thus, this review not only addresses a significant gap in the literature but also paves the way for future advancements in \ac{KC} diagnostics, offering valuable insights for researchers and healthcare practitioners.}

\subsection{Our Contributions}
While numerous review articles have examined the applications of \ac{TL} algorithms in medical imaging and disease detection—such as \citep{nasir2022kidney, kim2022transfer, liu2022development}—to the best of the authors' knowledge, no comprehensive survey has been conducted that specifically addresses the role of \ac{TL} in the context of kidney cancer (\ac{KC}). This paper aims to fill that gap by offering a novel and unified examination of how \ac{TL} and domain adaptation (\ac{DA}) techniques contribute to the detection and management of \ac{KC}.

Our work provides a structured and in-depth review of \ac{TL}'s applications across different types and stages of kidney cancer, integrating classification strategies and key conceptual frameworks. This unified perspective aims to demystify the role of \ac{TL} in \ac{KC} detection, with a particular focus on enhancements through fine-tuning and cross-domain \ac{DA}. Furthermore, this review underscores the importance of \ac{TL} in multi-source data fusion to improve diagnostic accuracy, and it highlights open challenges and future research directions, advocating for the development of more generalizable and computationally efficient \ac{TL} models for \ac{KC} detection. The unique contributions of this review are as follows:

\begin{itemize}
  \item Exploring various facets of \ac{TL} and \ac{DA} that have driven their evolution in medical imaging.
  \item Providing a detailed taxonomy of \ac{TL} and \ac{DA} frameworks, accompanied by an analysis of their foundational concepts.
  \item Conducting a comprehensive literature review on the role of \ac{TL} and \ac{DA} in \ac{KC} detection, based on consistent evaluation criteria. Demonstrating how \ac{TL} and \ac{DA} approaches can enhance integrated data pipelines for improved \ac{KC} diagnosis.
  \item Highlighting unresolved challenges in the application of \ac{TL} and \ac{DA} to cancer detection, including: (1) model generalizability and precision, (2) evaluation metric selection, (3) the risk of transferring task-irrelevant knowledge, among others.
  \item Proposing future research directions aimed at lowering computational complexity and resource demands in \ac{TL}-based frameworks.
  
\end{itemize}

Establishing this work as the first survey, to the best of the authors’ knowledge, to systematically address the integration of \ac{TL} and \ac{DA} within the \ac{KC} detection landscape. It contributes uniquely by (1) investigating diverse applications, (2) presenting an integrated framework, (3) identifying technical and methodological challenges, and (4) offering insights for future exploration. Table. \ref{table:1}, displays the results of the comparison of the contribution of the proposed review with other surveys in the field of \ac{TL}. {This study is intended for a diverse audience that includes researchers, healthcare professionals, and technical experts working in the fields of medical imaging, oncology, and {\ac{AI}}. It is particularly relevant for those interested in leveraging cutting-edge {\ac{DL}} and {\ac{TL}} techniques to improve early detection and diagnosis of {\ac{KC}}. This audience may consist of biomedical engineers, data scientists, and oncologists who are looking to enhance diagnostic accuracy, overcome challenges in dataset limitations, and explore innovative methodologies for precision medicine. Furthermore, the study is valuable for academics and industry professionals aiming to integrate advanced \ac{AI}-driven technologies into clinical practices for cancer diagnosis and treatment planning.}

\begin{table}[t!]
\caption{A summary of a comparison of the proposed study's contribution with other studies in the field of \ac{TL} is shown. A check mark (\cmark) signifies that the area has been covered, while a cross mark (\xmark) indicates that it has not been addressed.}
\label{table:1}
\small

\begin{tabular}{cccccccccccc}
\hline
Review & Year & Description & TLB & TLS & \ac{DA} & TLA & TLD & TLAD & TLL & KCH
& FD \\ \hline

\multicolumn{1}{l}{\citep{garcia2018mental}} &
\multicolumn{1}{l}{2018} & \multicolumn{1}{l}{\ac{TL} for mental health 
} & \cmark & \xmark & \xmark & \xmark & \cmark & \xmark & \xmark & \xmark & \xmark \\

\multicolumn{1}{l}{\citep{hu2018deep}} &
\multicolumn{1}{l}{2018} & \multicolumn{1}{l}{\ac{TL} for cancer detection } &
\cmark & \xmark & \xmark & \xmark & \cmark & \xmark & \xmark & \xmark & \cmark \\

\multicolumn{1}{l}{\citep{gardezi2019breast}} &
\multicolumn{1}{l}{2019} & \multicolumn{1}{l}{\ac{TL} for breast cancer} &
\cmark & \xmark & \xmark & \xmark & \cmark & \xmark & \xmark & \xmark & \xmark \\

\multicolumn{1}{l}{\citep{godasu2020transfer}} &
\multicolumn{1}{l}{2020} & \multicolumn{1}{l}{\ac{TL} for medical image} & \cmark & \xmark & \xmark & \xmark & \xmark & \xmark & \xmark & \cmark & \xmark \\

\multicolumn{1}{l}{\citep{wan2021review}} &
\multicolumn{1}{l}{2021} & \multicolumn{1}{l}{\ac{TL} for EEG signal analysis} & \cmark & \xmark & \xmark & \xmark & \xmark & \xmark & \xmark & \cmark & \xmark \\

\multicolumn{1}{l}{\citep{valverde2021transfer}} &
\multicolumn{1}{l}{2021} & \multicolumn{1}{l}{\ac{TL} for \ac{MRI} brain } &
\cmark & \xmark & \xmark & \xmark & \cmark & \xmark & \cmark & \xmark & \xmark \\

\multicolumn{1}{l}{\citep{agarwal2021transfer}} &
\multicolumn{1}{l}{2021} & \multicolumn{1}{l}{\ac{TL} for Alzheimer's disease} &
\cmark & \xmark & \xmark & \xmark & \xmark & \xmark & \xmark & \xmark & \xmark \\

\multicolumn{1}{l}{\citep{ardalan2022transfer}} &
\multicolumn{1}{l}{2022} & \multicolumn{1}{l}{\ac{TL} for neuroimaging analysis}
& \cmark & \xmark & \xmark & \xmark & \xmark & \xmark & \xmark & \cmark & \cmark \\

\multicolumn{1}{l}{\citep{abdelrahman2022kidney}} &
\multicolumn{1}{l}{2022} & \multicolumn{1}{l}{\ac{TL} for KC} &
\cmark & \cmark & \xmark & \xmark & \cmark & \xmark & \xmark & \xmark & \cmark \\

\multicolumn{1}{l}{\citep{cenggoro2023systematic}} &
\multicolumn{1}{l}{2023} & \multicolumn{1}{l}{\ac{TL} for COVID-19 medical image}
& \cmark & \xmark & \xmark & \xmark & \xmark & \xmark & \xmark & \xmark & \cmark \\

\multicolumn{1}{l}{\citep{bijam2023review}} &
\multicolumn{1}{l}{2023} & \multicolumn{1}{l}{\ac{TL} for Diabetic Retinopathy }
& \cmark & \xmark & \xmark & \xmark & \xmark & \xmark & \xmark & \xmark & \cmark \\

\multicolumn{1}{l}{Our} & \multicolumn{1}{l}{2024} & \multicolumn{1}{l}{\ac{TL}
for KC} & \cmark & \cmark & \cmark & \cmark & \cmark & \cmark & \cmark & \cmark & \cmark \\ 

\hline
\end{tabular}

\begin{flushleft}
Abbreviations: 
TL background (TLB),
TL segmentation (TLS),
TL applications (TLA),
TL datasets (TLD),
TL advantages (TLAD),
TL limitations (TLL),
Key challenges (KCH),
Futur direction (FD)
\end{flushleft}
\end{table}

\section{Methodology}\label{sec2}

\subsection{Search strategy and paper selection }

This survey on deep {\ac{TL}} for KC diagnosis provides a comprehensive exploration of existing {\ac{DL}}-based {\ac{TL}} frameworks and their application in the medical field, specifically targeting KC diagnosis. The methodological approach taken in this review encompasses several systematic steps to ensure a thorough analysis and presentation of the current state of research in this area.

The literature search was conducted using several academic databases including PubMed, IEEE Xplore, Scopus, Springer Nature, Elsevier, Wiley, Taylor \& Francis, ACM, and Google Scholar. The search was designed to identify relevant articles published in the last five years, ensuring the timeliness and relevance of the data collected. The keywords used for the search included combinations of ``{\ac{TL}}'', ``deep {\ac{TL}}'', ``{\ac{DA}}'', ``{\ac{KC}} diagnosis'', and ``medical imaging''. The search strategy was adapted to the syntax and capabilities of each database to maximize the retrieval of pertinent articles. The main inclusion and exclusion criteria, are:

\begin{itemize}
  \item \textbf{Inclusion criteria:}
    \begin{itemize}
      \item Studies that specifically apply {\ac{DL}} and {\ac{TL}} techniques.
      \item Articles focusing on the diagnosis of {\ac{KC}} using medical imaging.
      \item Papers published within the last five years to ensure current relevance.
      \item Studies providing clear methodological details and quantitative results.
    \end{itemize}
  \item \textbf{Exclusion criteria:}
    \begin{itemize}
      \item Non-English publications.
      \item Conference abstracts or presentations without full-text availability.
      \item Studies not involving explicit use of {\ac{DL}} or {\ac{TL}} frameworks.
      \item Papers not directly related to {\ac{KC}}.
    \end{itemize}
\end{itemize}

\subsection{Bibliometric Analysis}
\begin{figure*}[t!]
\begin{center}
\includegraphics[width=1\columnwidth]{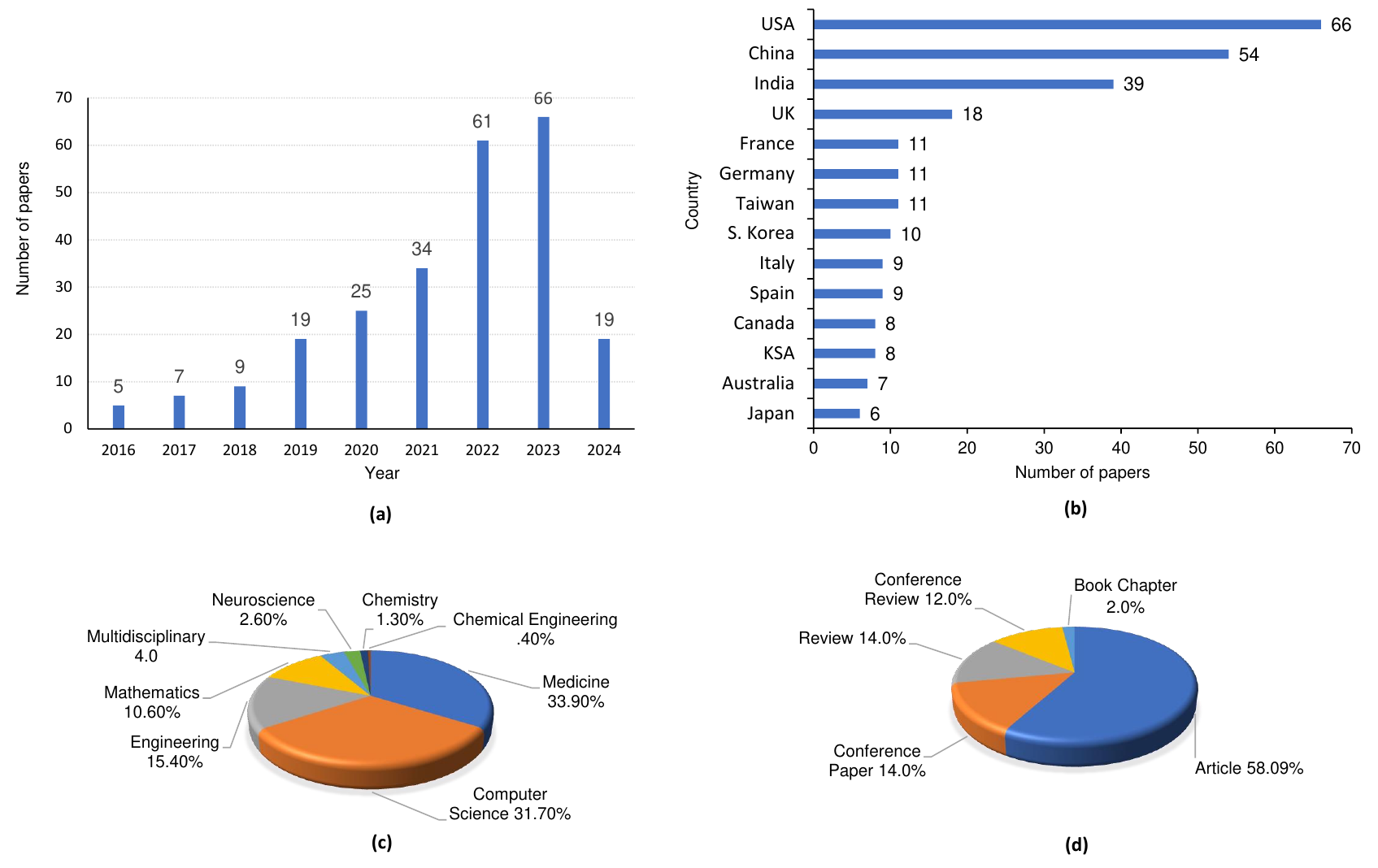}
\end{center}
\caption{{Bibliometric analysis in terms of: (a) The number of published articles; (b) The most country known by active researchers; (c) The percentage of the contribution in each domain; and (d) Distribution of paper types.}}
\label{fig3}
\end{figure*}

To explore and scrutinize the scientific investigations detailed in this review, bibliometric methods were utilized. The cumulative interest in \ac{TL}-based \ac{KC} over time is summarized in Figure  \ref{fig3}. This illustration provides a snapshot of the enduring interest in research on \ac{TL}-based \ac{KC}, highlighting a notable uptick in enthusiasm for crafting \ac{TL}-based \ac{KC} methodologies. Specifically, this surge in interest is represented in Figure  \ref{fig3}(a), indicating a sharp rise in the publication count, which peaked at 66 articles in 2023. Furthermore, Figure \ref{fig3}(b) displays the countries contributing most significantly to research outputs in this area, with China and the United States emerging as the most prolific in T\ac{TL}-based \ac{KC} research advancements. Furthermore, Figure  \ref{fig3}(c) identifies the differents domain of application of \ac{KC}-oriented \ac{TL}. Lastly, Figure  \ref{fig3}(d) delineates the distribution of paper types, revealing that journal articles predominate, constituting 65.8\% of all published work, with conference papers making up 19.7\%.

Initial screening was based on titles and abstracts, with articles potentially meeting the inclusion criteria selected for full-text review. This stage filtered out irrelevant or out-of-scope studies, streamlining the focus on {\ac{TL}} in the context of {\ac{KC}}. 
Essential information was extracted from each article, including study objectives, {\ac{DL}} models, datasets, {\ac{TL}} techniques employed, and principal findings. This structured approach facilitated comparative analysis and synthesis.

The structure of this document is outlined below. Section~2 describes the methodology used in conducting this review, including the search strategy and bibliometric analysis. Section~3 presents a comprehensive taxonomy of \ac{TL} and domain adaptation (\ac{DA}) techniques applied in the context of \ac{KC}, including inductive, transductive, domain generalization, self-taught, and few-shot learning, along with associated strategies such as fine-tuning and multi-task learning. Section~4 explores key applications of \ac{TL} in \ac{KC}, encompassing cancer detection, \ac{CT} scan analysis and segmentation, genomic data analysis, multi-modal data fusion, and explainable AI. Section~5 provides an overview of the most commonly used datasets and evaluation metrics in this domain. Challenges encountered in employing \ac{TL} for \ac{KC} diagnosis and potential directions for future research are discussed in Sections~6 and~7, respectively. Finally, Section~8 summarizes the conclusions drawn from this comprehensive review.

\section{Taxonomy of TL}\label{sec3}
\ac{TL} is a method in \ac{ML} that involves applying knowledge obtained from solving one problem to a different, yet related, problem. This part presents the mathematical framework underlying \ac{TL}, beginning with the concepts of a "domain" and a "task."

\noindent A \textbf{domain} \( \mathcal{D} \) consists of two primary elements:

\begin{itemize}
\item A feature space \( \mathcal{X} \), encompassing all potential inputs.
\item A marginal probability distribution \( P(X) \), with \( X \in \mathcal{X} \), which describes the distribution of samples across the feature space.
\end{itemize}

\noindent A \textbf{task} \( \mathcal{T} \) is distinguished by: A label space $\mathcal{Y}$; A goal-oriented predictive function $f(\cdot)$   that associates an input \( x \in \mathcal{X} \) with an output  $y \in \mathcal{Y}$. This function  $f$  is generally unknown and represents the target of learning algorithms to estimate.

In \ac{TL}, we deal with a source domain \( \mathcal{D}_S \) and a target domain \( \mathcal{D}_T \), along with their corresponding tasks \( \mathcal{T}_S \) and \( \mathcal{T}_T \). The objective is to improve the learning of the target predictive function \( f_T(\cdot) \) in \( \mathcal{D}_T \) using the knowledge from \( \mathcal{D}_S \) and \( \mathcal{T}_S \).

In the context of \ac{TL}, we engage with both a source domain \( \mathcal{D}_S \) and a target domain \( \mathcal{D}_T \), as well as their respective tasks \( \mathcal{T}_S \) and \( \mathcal{T}_T \). The aim is to enhance the learning of the target predictive function \( f_T(\cdot) \) within \( \mathcal{D}_T \) by leveraging the knowledge from \( \mathcal{D}_S \) and \( \mathcal{T}_S \).

The purpose of a \ac{TL} approach is to utilize the source domain and task to improve the accuracy of the predictive function in the target domain. This involves adapting models, identifying shared feature representations, or employing other methods to make use of the similarities between the domains and tasks.  Table \ref{table:3}, summarizes the types of \ac{TL}. Table \ref{table:4} provides a summary of various cutting-edge techniques that utilize \ac{TL} types for \ac{KC}, including the datasets and metrics employed.

\begin{table}[t]
\caption{Types of \ac{TL}.}
\label{table:3}
\begin{tabular}{cccc}
\hline
\ac{TL} settings & Source & Target & Tasks\\
\hline

\multicolumn{1}{l}{Traditional \ac{ML}/\ac{DL}} & $D_s=D_T$ & $D_s=D_T$ & \multicolumn{1}{l}{All}\\

\multicolumn{1}{l}{Inductive \ac{TL}} & $D_s \cong D_T$ & $D_s \neq D_T$ & \multicolumn{1}{l}{Classification and regression}\\

\multicolumn{1}{l}{Transductive \ac{TL}} & $D_s \cong D_T$ & $D_s=D_T$ & \multicolumn{1}{l}{Classification and regression}\\

\multicolumn{1}{l}{Unsupervised \ac{TL}} & $D_s \neq D_T$ & $D_s \neq D_T$ & \multicolumn{1}{l}{Clusterings and reduction}\\

\hline
\end{tabular}%
\end{table}

\subsection{Inductive \ac{TL}}
Inductive \ac{TL} focuses on using the knowledge gained from the source domain and task to improve learning in a related but different target task. The key elements are defined as follows:
\begin{itemize}
    \item Let $\mathcal{D}_S = \{X_S, P(X_S)\}$ and $\mathcal{D}_T = \{X_T, P(X_T)\}$ be the source and target domains, respectively, where $X$ represents the input space and $P(X)$ the marginal probability distribution.
    \item The tasks $\mathcal{T}_S = \{Y_S, f_S(\cdot)\}$ and $\mathcal{T}_T = \{Y_T, f_T(\cdot)\}$, with $Y$ being the output space and $f(\cdot)$ the target predictive function.
    \item The goal is to improve $f_T$ by utilizing $f_S$, often requiring some labeled data in $\mathcal{D}_T$.
\end{itemize}

\subsubsection{Fine-tuning}
\begin{figure*}[t]
\begin{center}
\includegraphics[width=1\columnwidth]{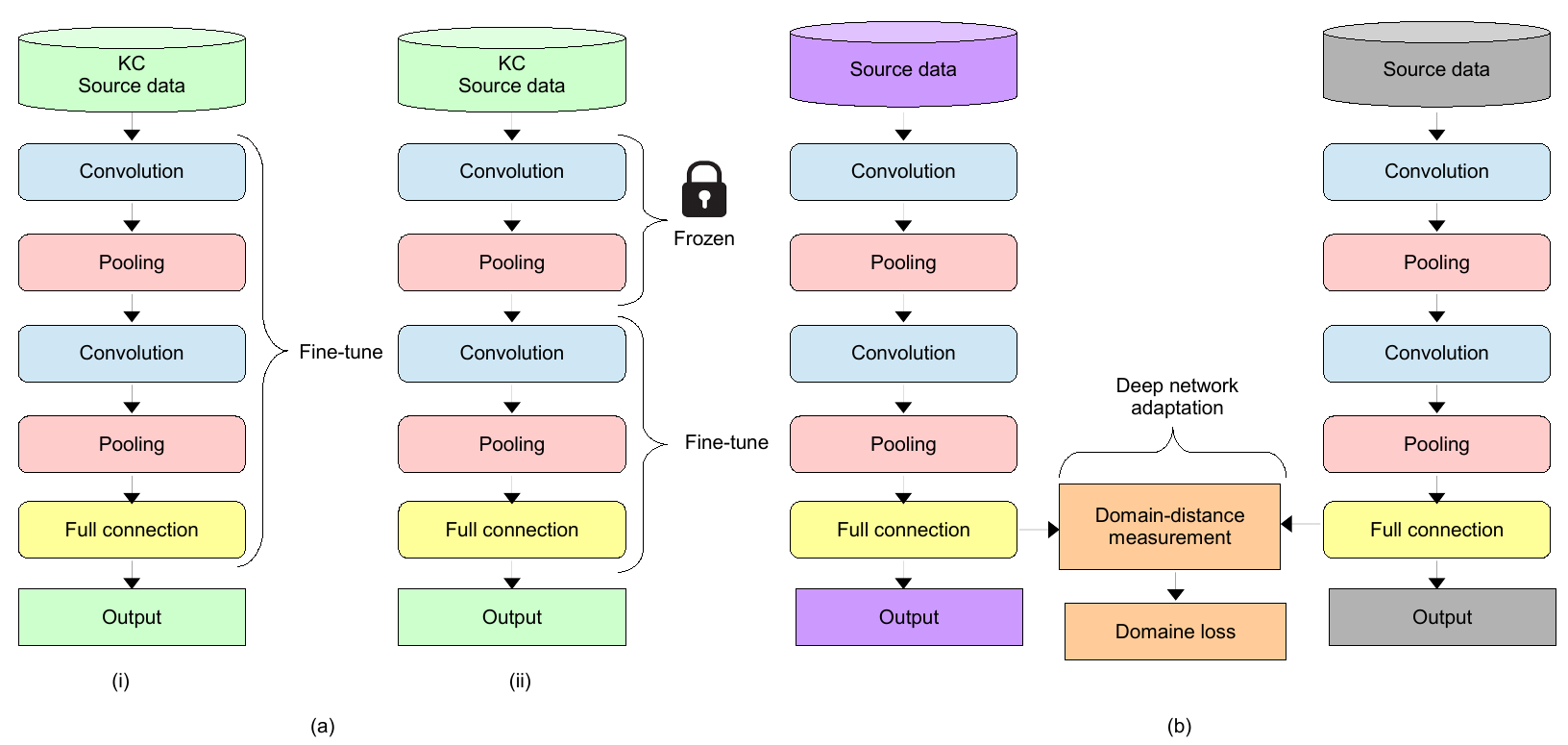}
\end{center}
\caption{Example of \ac{TL} models used in KC: (a) Fine-tuning, and (b) DA.}
\label{fig4}
\end{figure*}

During fine-tuning, the model's weights, initially learned from the source domain, are adjusted to better suit the target task (Figure \ref{fig4}(a)). This adjustment is done by continuing the training process on the target dataset, often with a smaller learning rate to make subtle changes to the weights. Fine-tuning in this context allows the model to adapt the knowledge acquired from the source domain to the specifics of the target domain and task, making it a quintessential example of inductive \ac{TL} \cite{lachenani2024improving,djeffal2024transfer,habchi2024ultrasound}. Figure \ref{fig4}(a) illustrates two methods of fine-tuning \ac{CNN}: (i) to fine-tune the weights of the whole network and (ii) to fine-tune some certain layers. Figure \ref{fig4}(b) showcases a \ac{CNN} instance of deep network adaptation, which modifies the distribution in the full connection layer by measuring the domain distance. {For example in \citep{asif2022modeling}, the authors' research highlights the significance of fine-tuning in \ac{TL} for enhancing a pre-trained VGG19 and naïve inception model's ability to detect kidney diseases from \ac{CT} images. By adapting the \ac{DL} model more specifically to this task, fine-tuning significantly boosts performance, achieving an impressive accuracy in identifying various kidney diseases. Similarly, the article \cite{rana2024kidneymultinet} presents KidneyMultiNet, a web-based system for kidney disease detection using a hybrid \ac{ML} model that combines two \ac{CNN}s, DenseNet201 and Xception, and leverages \ac{TL} techniques. This system, trained on a publicly available kidney CT scan dataset, achieved high performance and demonstrated the growing potential of advanced machine learning techniques in medical diagnostics. Furthermore, \citep{matos2023evaluation} introduces a methodology for kidney segmentation using \ac{TL} with the U-Net architecture, employing a pre-trained EfficientNet as an encoder, which results in impressive segmentation precision and contributes to enhanced kidney tumor detection and treatment planning. Likewise, Kalluraya Puttur et al. \citep{puttur2024advanced} introduced a \ac{DL}-based system for early \ac{KC} detection using \ac{CT} scans, building on the VGG16 architecture and enhanced with \ac{TL}. This approach demonstrated the potential of \ac{DL} in improving diagnostic accuracy and patient outcomes. In addition, the study \citep{krishnan2024deep} explores the transfer of a model trained on human \ac{MRI} of ADPKD to animal models, effectively addressing challenges such as class imbalance and small datasets, highlighting the effectiveness of \ac{TL} in cross-species medical imaging applications. Similarly, the study \citep{liu2024renal} introduces a \ac{TL} model combining contrastive learning and ResNet-50 for renal pathology image classification, demonstrating how contrastive learning and \ac{TL} can improve performance with minimal labeled data.  In parallel, Sharma et al. \citep{sharma2024transfer} present an \ac{AI}-based system using a modified InceptionV3 model for diagnosing kidney abnormalities, offering an autonomous and accurate diagnostic tool, particularly beneficial in areas with limited access to renal specialists. Additionally, Kaur et al. \citep{kaur2024advancements} propose a novel \ac{DL} model based on EfficientNetB3 for classifying kidney diseases into four categories, demonstrating its high accuracy and practical utility in early diagnosis and treatment planning. In a similar vein, Prasher et al. \citep{prasher2024vgg16} introduce a VGG16 \ac{TL} model for diagnosing kidney stones using coronal \ac{CT} images, highlighting its potential for clinical application by reducing dependence on manual diagnosis. Meanwhile, the study \citep{nandhitha2024impact} investigates the role of activation functions in SqueezeNet’s performance for kidney disease classification, revealing that ReLU provides the best results, thereby emphasizing the influence of activation functions on medical image classification outcomes.  Ghosh et al. \citep{ghosh2025fuzzy} propose STREAMLINERS, a model that integrates fuzzy logic and \ac{TL}-based feature extraction using DenseNet121 and ResNet101, achieving high accuracy on enhanced datasets. Similarly, Kumar et al. \citep{kumar2024enhancing} develop a VGG16-based model that outperforms several alternatives in classifying kidney diseases, enhancing accuracy through fine-tuning and feature extraction, though it requires large labeled datasets. The study \citep{goker2024transfer} proposes an image processing model for kidney stone classification using \ac{CT} images, with image enhancement techniques like CLAHE and Laplacian filtering, and finds that ResNet101 performs best. This approach provides a reliable diagnostic tool for timely kidney stone detection.  In the study \citep{aljaaf2018early}, the authors introduce a smart model for kidney disease prediction, incorporating fine-tuning via a contrastive version of the wake-sleep algorithm, which effectively slows down kidney damage progression. Similarly, \citep{kang2024computed} presents a \ac{DL}-driven segmentation framework that emphasizes the role of fine-tuning and novel loss functions to enhance model accuracy, particularly in kidney and renal tumor segmentation. Lastly, in \citep{yang20223d}, fine-tuning is used to adapt a 3D \ac{CNN} (NephNet3D) to classify cell types in human kidney tissue, resulting in significant improvements in predictive performance.}

\subsubsection{Multi-Task learning (MTL)}
Multi-task Learning aims at simultaneously learning multiple related tasks to improve generalization. It is mathematically described as:
\begin{itemize}
    \item Given a set of tasks $\{\mathcal{T}_1, \mathcal{T}_2, \ldots, \mathcal{T}_N\}$, the objective is to learn these tasks jointly by exploiting commonalities and differences.
    \item This involves optimizing a shared representation $\Phi$ and task-specific functions \\ $\{f_{T_1}, f_{T_2}, \ldots, f_{T_N}\}$.
    \item The objective function is
    
    $\min_{\Phi, f_{T_1}, \ldots, f_{T_N}} \sum_{i=1}^{N} \alpha_i  \mathcal{L}_{T_i}(f_{T_i}(\Phi(X_{T_i})), Y_{T_i})$, where $\mathcal{L}_{T_i}$ is the loss function for task $i$.
\end{itemize}

The authors' objective in \citep{bernardini2021semi} is to introduce an innovative semi-supervised multi-task learning method designed for forecasting the short-term progression of kidney disease using data from general practitioners' electronic health records. They have shown that their methodology can: (i) track the progression of estimated glomerular filtration rate over time by enforcing a connection between sequential time periods, and (ii) harness valuable insights from patient data without labels, particularly when the data with labels are in limited supply.

In \citep{pan2021multi}, the authors introduced a new multi-task learning approach aimed at assessing image quality, performing de-blurring, and classifying kidney diseases. The study analyzed immunofluorescence images from 1,608 patients, dividing them into 1,289 for training and 319 for testing. This innovative technique was not only effective in diagnosing four different kidney diseases using blurred immunofluorescence images, but it also demonstrated strong capabilities in two additional tasks: evaluating image quality and removing blur from images.

In \citep{keshwani2018computation}, the authors introduce a multi-task 3D \ac{CNN} designed for segmenting autosomal dominant polycystic kidney disease, which is identified by the continuous enlargement of renal cysts and stands as the most common severe hereditary kidney disorder. They successfully attained a high average dice score through this approach.

\subsection{Transductive \ac{TL}}
The objective of transductive \ac{TL} is to predict the labels 
$\{y_j^T\}_{j=1}^{n_T}$ for the instances in the target domain $\mathcal{D}_T$, leveraging the knowledge from the source domain $\mathcal{D}_S$, where:

\begin{itemize}
    \item A source domain $\mathcal{D}_S = \{(\mathbf{x}_i^S, y_i^S)\}_{i=1}^{n_S}$ with $n_S$ labeled instances, where $\mathbf{x}_i^S \in \mathbb{R}^d$ represents the feature vector of the $i$-th instance and $y_i^S \in \mathcal{Y}$ represents its label.
    \item A target domain $\mathcal{D}_T = \{\mathbf{x}_j^T\}_{j=1}^{n_T}$ with $n_T$ unlabeled instances, where $\mathbf{x}_j^T \in \mathbb{R}^d$.
\end{itemize}

Transductive \ac{TL} can be further categorized based on the strategies employed to bridge the gap between the source and target domains:

\subsubsection{Domain adaptation}
It focuses on adapting the source domain model to work effectively on the target domain by minimizing the domain shift.  In other word, the aim of \ac{DA} is to minimize the domain shift between the source domain $\mathcal{D}_S$ and the target domain $\mathcal{D}_T$ (Figure \ref{fig4}(b)) \citep{farahani2021brief}. Techniques include: 

\begin{itemize}
\item \emph{Feature representation alignment:} Aim to find a transformation $\phi$ that aligns the feature distributions of the source and target domains by minimizing a distance metric $d(\phi(\mathcal{D}_S), \phi(\mathcal{D}_T))$, such as the maximum mean discrepancy \citep{chen2019progressive}.
    
\item \emph{Domain-invariant feature learning:} Learn a feature representation $\phi$ that is invariant to the change in domains, optimizing $\phi$ such that the predictive model $f(\phi(\cdot))$ achieves low error on both $\mathcal{D}_S$ and $\mathcal{D}_T$ \citep{li2018domain}.
    
\item \emph{Discrepancy minimization:} Directly minimize the discrepancy between the source and target domain distributions using techniques like adversarial training, where a discriminator is trained to distinguish between the source and target domain features while the feature extractor $\phi$ is trained to fool the discriminator \citep{du2021cross}.
\end{itemize}

Instance weighting, feature transformation, and kernel methods are key techniques used in the \ac{DA} process, aimed at addressing the issue of domain shift—where the data distribution in the source domain differs from that in the target domain \citep{ishii2017joint}. These techniques facilitate the transfer of knowledge from a source domain to a target domain, helping to improve the performance of \ac{ML} models on the target domain despite differences in data distribution \citep{zhang2024double}. 

\begin{itemize}
\item \emph{Instance weighting:} It assigns weights to instances in the source domain based on their relevance to the target domain. The goal is to give higher importance to source instances that are more similar to the target domain, thereby reducing the bias towards the source domain distribution \citep{ting2002instance}. Typically, given a source instance $(x_i^S, y_i^S)$, assign a weight $w_i$ based on its similarity to the target domain. The weighted loss for the source domain is then $\sum_{i=1}^{n_S} w_i \ell(f(x_i^S), y_i^S)$, where $\ell$ is a loss function \citep{wang2017instance}.

\item \emph{Feature transformation:} It transforms the feature space to find a common subspace where the source and target domain data can be directly compared or combined. Methods include: 

\begin{itemize}
    \item \emph{Subspace alignment:} Find transformations $\phi_S$ and $\phi_T$ for the source and target domains, respectively, that align their subspaces, minimizing $\|\phi_S(\mathcal{D}_S) - \phi_T(\mathcal{D}_T)\|$ \citep{fernando2014subspace}.
    
    \item \emph{Manifold learning:} Discover a low-dimensional manifold that both domains can be projected onto, learning a mapping $\phi$ that captures the intrinsic geometry of the data \citep{wang2018visual}.
    
    \item \emph{DL approaches:} Use neural networks to learn a domain-agnostic feature representation, possibly with domain-adversarial components \citep{cai2019unsupervised}.
\end{itemize}

\item \emph{Kernel methods:}  They employ kernel functions to map the original feature space into a higher-dimensional space, where the source and target domains are more comparable.
This approach facilitates finding a common representation that reduces the distributional differences between the domains \citep{gong2017geodesic}. Given a kernel function $k(\cdot, \cdot)$, transform instances $x_i^S$ and $x_j^T$ into a higher-dimensional space via $\phi$, where $\phi$ is implicitly defined by $k$. The goal is to learn a predictive model in this space that reduces the domain discrepancy \citep{wang2019semi}.
\end{itemize}
\vspace{0.3cm}
These strategies are not mutually exclusive and can be combined to develop more sophisticated transductive \ac{TL} models that effectively utilize the knowledge from the source domain to make accurate predictions in the target domain. {The study \citep{iqbal2025novel} presents reciprocal domain adaptation network for \ac{KC} diagnosis, which integrates \ac{TL} with mutual model adaptation and domain model learning. While mutual model adaptation extracts rich features through global and local pyramid pooling, domain model learning enhances representation learning by combining semi-supervised, domain-independent features. Consequently, reciprocal domain adaptation network effectively addresses issues like data bias, domain shifts, and privacy, providing a promising solution for early KC detection. In a similar context, Canbay et al. \citep{canbay2024privacy} propose a privacy-preserving \ac{TL} framework, incorporating differential privacy with \ac{CNN}s. Although classification accuracy decreases with increased privacy settings, the framework ensures strong performance while protecting patient data, thus offering a secure solution for early kidney disease detection. Furthermore, the study by Kwon et al. \citep{kwon2024transfer} focuses on improving segmentation for autosomal dominant polycystic kidney disease using U-Net and \ac{TL} with \ac{MRI} data. By testing different training scenarios, the study demonstrates that single-shot fast spin-echo images yield the best results, reinforcing the utility of \ac{TL} in medical segmentation. Similarly, Lopez et al. \citep{lopez2023boosting} present a two-step \ac{TL} method using ResNet50 for classifying kidney stones in endoscopic and charge-coupled device images. This approach improves accuracy and outperforms traditional models. In \citep{hong2022source}, the authors propose a two-stage source-free unsupervised domain-adaptation framework for cross-modality abdominal multi-organ segmentations. The proposed method only requires a well-trained source model and an unlabeled target dataset, and does not need access to the source data. The experimental results highlight that this approach achieves satisfactory performance in adapting a labeled \ac{CT} dataset to an unlabeled \ac{MRI} dataset for abdominal organ segmentations, including the right and left kidneys. Additionally, in \citep{wang2023rethinking}, the authors present a new \ac{DA} approach called dimension-based disentangled dilated \ac{DA}, designed to tackle the problem of domain shift for medical image segmentation tasks without the annotations of the target domain. By using adaptive instance normalization, the method encourages the content information to be stored in the spatial dimension and the style information in the channel dimension. They validate the proposed method for cross-modality medical image segmentation tasks on public datasets, including liver, right kidney, left kidney, and spleen.}

\subsubsection{Domain generalization}
{\Ac{DG}} focuses on the challenge of developing models that can generalize well to unseen domains. In {\ac{DG}}, a model is trained on data from multiple source domains to perform effectively on data from target domains that were not encountered during training. The core objective is to learn a domain-invariant representation that captures the underlying structure of the data, enabling the model to make accurate predictions across different, unseen domains \citep{ghifary2016scatter}. 
In {\ac{DG}}, the goal is to learn a model $f$ that generalizes well to any target domain $D_T$, given multiple source domains $\{D_1, D_2, \ldots, D_K\}$ \citep{zhou2022domain}:

\begin{equation}
\min_f \sum_{k=1}^K \mathcal{L}(f(X_k), Y_k) + \lambda \Omega(f)
\end{equation}

In \citep{kline2021improving}, the authors present an innovative approach utilizing neural style transfer to reduce model bias towards texture and intensity, aiming to prioritize shape characteristics. They conduct experiments using 200 T2-weighted MR images, half with fat-saturation and half without, to segment kidneys in patients with polycystic kidney disease. The results demonstrate a significant improvement in segmentation performance, indicating better alignment with actual kidney shapes. 

In \citep{su2023rethinking}, the authors address the challenge of single-source {\ac{DG}} in medical image segmentation, specifically noting the issue of domain shifts common in clinical image datasets. Traditional approaches, relying on global or random data augmentation, often produce samples that lack diversity and informativeness, inadequately representing the potential target domain distribution. To tackle this, the authors propose a novel data augmentation strategy for {\ac{DG}}, inspired by the inherent class-level representation invariance and style mutability of medical images. They suggest that unseen target data could be represented as a linear combination of class-specific random variables, each adhering to a location-scale distribution. This concept is operationalized through the use of constrained bézier transformation for augmenting data at both global and class-specific levels, significantly enhancing augmentation diversity. Additionally, a saliency-balancing fusion mechanism is introduced to increase the informativeness of augmented samples by utilizing gradient information for orientation and magnitude guidance. The authors theoretically demonstrate that their augmentation strategy effectively limits the generalization risk for unseen target domains. Their saliency-balancing location-scale augmentation method notably outperforms existing state-of-the-art techniques in two single-source {\ac{DG}} tasks, underscoring the potential of their approach to improve kidney and other medical image segmentation tasks by addressing the challenges posed by domain shifts.

\subsubsection{Self-Taught learning}
In self-taught learning, the model learns from unlabeled data that is not necessarily from the same distribution as the labeled data it was initially trained on. The key idea behind self-taught learning is to leverage large amounts of unlabeled data to improve the learning algorithm's performance on a related task for which labeled data is scarce \citep{michieli2020adversarial}. The process typically involves two main steps: first, extracting useful features or representations from the unlabeled data, and then, using these features to enhance the learning process on the labeled dataset. Self-taught learning enables the model to generalize better to new, unseen data by exploiting the underlying structure shared between the labeled and unlabeled datasets, even when they come from different domains. This approach is particularly valuable in scenarios where acquiring labeled data is expensive or labor-intensive, allowing for more efficient use of available data resources \citep{zhang2023cross}.

\begin{equation}
\min_f \mathcal{L}(f(X_S), Y_S) + \alpha \mathcal{L}_{AE}(X_T)
\end{equation}

\subsubsection{Few-Shot learning}
{\Ac{FSL}} is a subset of transductive \ac{TL} that focuses on the ability of a model to learn and make accurate predictions from a very limited number of training examples. The primary challenge in {\ac{FSL}} is to design algorithms that can quickly adapt to new tasks or classes using only a few labeled samples, often referred to as "shots." This approach is crucial in scenarios where collecting large amounts of labeled data is impractical or impossible, such as in medical diagnosis, where rare diseases may only have a handful of known cases. {\ac{FSL}} leverages prior knowledge, gained from related tasks or a broader dataset, to infer the new task's structure with minimal data. Typically, In {\ac{FSL}}, the model learns from a few examples in the target domain through a meta-learning approach:

\begin{equation}
\min_f \sum_{\tau \in T} \mathcal{L}(f_{\theta + \Delta \theta_\tau}(X_\tau), Y_\tau)
\end{equation}

This involves techniques such as meta-learning, where the model learns learning strategies from multiple tasks, or embedding learning, where data is transformed into a space where tasks with little data can be more easily learned. {\ac{FSL}} aims to mimic human learning's efficiency and flexibility, enabling models to generalize well from limited information.

In \citep{mendez2022generalization}, the authors address the challenge of {\ac{DA}} in \ac{DL} for medical imaging, specifically in classifying kidney stones from endoscopic images taken under various conditions. They introduce a novel approach combining {\ac{FSL}} with meta-learning. Initially, a self-supervised learning phase pre-trains the model to improve feature generalization across domains. Subsequently, meta-learning fine-tunes these features specifically for the domain of kidney stones, adapting the model to new, unseen domains effectively. This method significantly enhances the model's generalization ability and demonstrates a promising direction for overcoming {\ac{DA}} challenges in medical imaging applications.

In this study \citep{kim2021bidirectional}, the authors present a 3D few-shot segmentation framework tailored for precise segmentation of conditions such as \ac{KC}, despite the scarcity of target organ annotation in training samples. A network resembling U-Net is developed to infer segmentation by understanding the relationship between 2D slices from support data and a query image, incorporating a bidirectional gated recurrent unit to ensure consistency in the encoded features among adjacent slices. Furthermore, they propose a \ac{TL} approach to adjust the model to the specificities of the target image and organ by pre-testing updates using arbitrarily selected support and query samples. This model significantly outperforms existing few-shot segmentation models and rivals fully supervised models trained on more extensive target data.

\begin{table*}[ht!]
\centering
\caption{A summary of \ac{TL} types for \ac{KC}.}
\label{table:4}
\scriptsize
\begin{tabular}{
    >{\centering\arraybackslash}m{0.5cm} 
    >{\centering\arraybackslash}m{0.4cm} 
    >{\centering\arraybackslash}m{2cm} 
    >{\centering\arraybackslash}m{1.1cm} 
    >{\centering\arraybackslash}m{1.5cm} 
    >{\centering\arraybackslash}m{1.5cm} 
    >{\centering\arraybackslash}m{1.5cm} 
    >{\centering\arraybackslash}m{2cm} 
    >{\centering\arraybackslash}m{2cm} 
}
\hline
Ref& {Year} & Model-based & \ac{TL} & \ac{TL} type & KC task & Datasets & Image features & Metrics  \\ \hline

\citep{asif2022modeling} & {2022} &
VGG19 and Naïve inception & Inductive & Fine-tuning & Detection & Private data & 4000, CT & ACC  \\ \hline

\citep{statkevych2022improving} & {2022} & U-Net & Inductive & Fine-tuning & Segmentation & HuBMAP & 15, CT & Dice score  \\ \hline

\citep{kang2024computed} & {2024} & U-Net & Inductive & Fine-tuning & Segmentation & Private data & CT & -  \\ \hline

\citep{yang20223d} & {2023} & 3D-MS-RFCNN & Inductive & Fine-tuning & Segmentation & KiTS & 1059, 512×512, CT & ACC and dice score  \\ \hline

\citep{pan2021multi} & {2021} & CNN & Inductive & MTL & Classification & Private data &1608, CT & ACC and AUC  \\ \hline

\citep{keshwani2018computation} & {2018} & 3D CNN & Inductive & MTL & Segmentation & Private data &203, CT & Dice score  \\ \hline

\citep{hong2022source} & {2022} & U-Net & Transductive & DA & Segmentation & Private data &30, CT and 20, MRI & Dice score  \\ \hline

\citep{kline2021improving} & {2021} & CNN & Transductive & DG & Segmentation & Private data &200, MRI & Dice score  \\ \hline

\citep{su2023rethinking} & {2023} & U-Net with EfficientNet-b2 & Transductive & DG & Segmentation & Private data & MRI & Dice score  \\ \hline

\citep{mendez2022generalization} & {2022} & DL & Transductive & FSL & Classification & Private data &600, CT & ACC  \\ \hline

\citep{ma2019affinitynet} & {2019} & KNN & Transductive & FSL & Prediction & GDCD &  654 samples & ACC  \\ \hline

\citep{kim2021bidirectional} & {2021} & U-Net & Transductive & FSL & Segmentation & BCV &  30, 3D, CT & Dice score  \\ \hline

\citep{sun2022few} & {2022} & CNN & Transductive & FSL & Segmentation & Private data &  3D, 20, MRI, 30, CT & Dice score  \\ \hline

\hline
\end{tabular}
\end{table*}

\section{Applications of \ac{TL} in \ac{KC}}\label{sec4}
\ac{TL} has been applied in the field of \ac{KC} diagnosis and prognosis using medical imaging and genomic data. { Figure \ref{fig5} illustrate the applications of \ac{TL} in \ac{KC}. Table. \ref{table:2}, presents a summary of \ac{TL} research conducted in the field of kidney disease diagnosis.}  Some applications include:

\begin{figure*}[ht!]
\begin{center}
\includegraphics[scale=0.7]{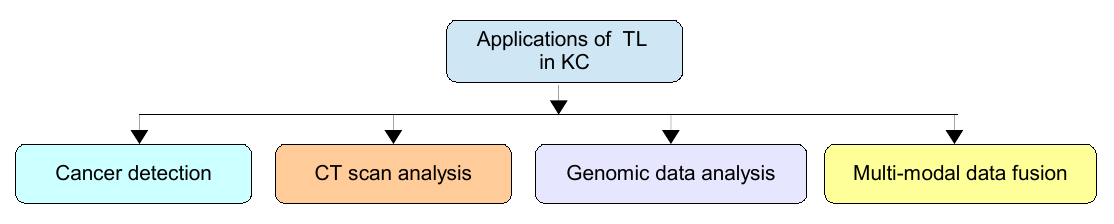}
\end{center}
\caption{The \ac{TL}-based applications of  for KC.}
\label{fig5}
\end{figure*}

\subsection{Cancer detection \ac{TL}} 
\begin{figure*}[ht!]
\begin{center}
\includegraphics[width=0.9\columnwidth]{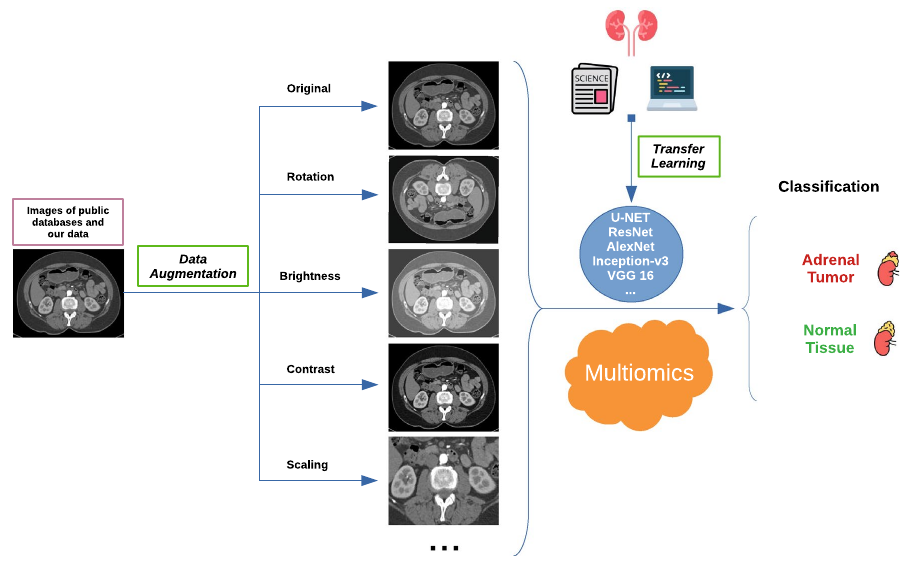}
\end{center}
\caption{{An example that shows a workflow for classifying renal tumors using \ac{DL} techniques. It starts with data augmentation, where images from public databases and internal data are processed through transformations like rotation, brightness adjustment, contrast enhancement, and scaling. These augmented images are then fed into various \ac{DL} models, such as U-Net, ResNet, AlexNet, Inception-v3, and VGG-16. \ac{TL} is applied to improve the model’s accuracy and reduce training time. The models then classify the images into categories: Adrenal Tumor or Normal Tissue. Additionally, multiomics data is integrated to enhance classification accuracy, incorporating both imaging and molecular data \cite{amador2024deep}.}}
\label{fig6}
\end{figure*}

\ac{TL} is a method in \ac{DL} that utilizes knowledge acquired from one task to enhance performance on a novel, but related, task, as outlined in \citep{sohail4348272deep}. Within the scope of analyzing medical images, \ac{TL} enables the training of models on extensive datasets of analogous images, like radiology scans, before refining these models on a smaller set of medical images designated for a specific purpose, such as the detection of \ac{KC}. To deploy \ac{TL} for \ac{KC} detection, the initial step involves gathering a comprehensive dataset of medical scans, such as \ac{CT} or \ac{MRI} images, which includes both healthy kidneys and those with cancerous growths. This collection serves as the foundation for pre-training a \ac{CNN} to differentiate between normal and cancerous images. After the \ac{CNN} has been pre-trained on the extensive dataset, it is then fine-tuned using a more focused collection of kidney images, precisely for \ac{KC} detection. This fine-tuning adjusts the \ac{CNN}'s pre-trained weights to refine its performance for this specific detection task. Throughout the fine-tuning phase, the \ac{CNN} hones in on distinct characteristics within the kidney images that signal the presence of cancerous tumors, as illustrated in Figure \ref{fig6}. This process benefits from the insights gained from the larger dataset of similar images, allowing the \ac{CNN} to learn how to identify \ac{KC} in the smaller dataset with increased speed and accuracy \citep{himeur2022using}.

\subsection{\ac{CT} scan analysis, and segmentation \ac{TL}} 
\begin{figure}[ht!]
\begin{center}
\includegraphics[width=0.7\columnwidth]{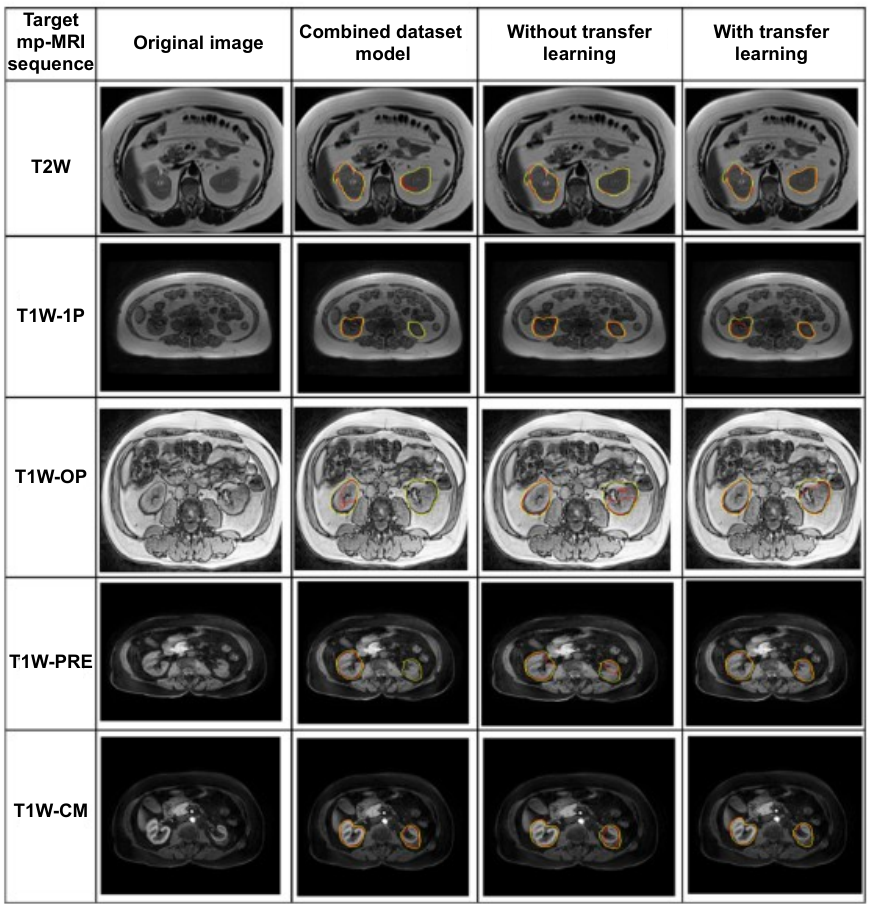}
\end{center}
\caption{{Example of the  application of \ac{TL} and segmentation in kidney cancer prognosis. It presents kidney segmentation on axial slices using a combined dataset model's predictions in column 2, the model predictions without \ac{TL} in column 3, and those with \ac{TL} in column 4. Each row corresponds to a distinct test case, where yellow contours represent the ground truth kidney labels, and red contours indicate the model's predictions
 \cite{gaikar2022transfer}.}}
\label{fig7}
\end{figure}

\ac{TL} is a prominent strategy in \ac{DL}, facilitating the adaptation of pre-trained models to novel tasks. This method has been particularly effective in the medical imaging domain for tasks such as the automatic segmentation applications (Figure \ref{fig7}) and classification of kidney tumors from \ac{CT} scans. \ac{TL} operates on the principle of leveraging the insights acquired from a model trained on a related issue trained on a similar dataset. This strategy is advantageous because it conserves time and resources that would otherwise be expended on initiating training from the ground up, while also enhancing accuracy through the transference of knowledge from the pre-trained model. Specifically, in the analysis of \ac{CT} scans, \ac{TL} has proven beneficial in refining pre-trained \ac{DL} models for more precise and efficient identification and categorization of kidney tumors.  {Additionally, Gido et al. \citep{gido2025kidney} present a \ac{DL} approach for kidney and tumor segmentation using 3D \ac{CT} data, demonstrating that fine-tuning is optimal for kidney segmentation, while mixed datasets are more effective for tumor segmentation. The authors in \citep{statkevych2022improving} emphasize the importance of hyperparameter fine-tuning, test time and train time data augmentations, and normalization in improving model performance for medical image segmentation tasks. The paper \citep{nagarajan2024ensemble} introduces an ensemble \ac{TL}-based model for kidney segmentation, incorporating multiple pre-trained \ac{CNN}s and advanced data augmentation to enhance segmentation robustness and generalization. Similarly, Andreini et al. \citep{andreini2024enhancing} adopt a cross-species \ac{TL} strategy using MobileNet and SegNeXt for glomeruli segmentation, achieving strong performance even with limited annotations. Moreover, the study by Miao et al. \citep{miao2024development} develops a CNN with \ac{TL} (ResNet-18) for segmenting ultrasound images of renal tissue ablation, outshining conventional methods and performing comparably to manual segmentation by experts.}

{\fontsize{6.5}{7.5}\selectfont
\begin{longtable}{
m{0.5cm} 
m{1.5cm} 
m{2cm} 
m{6cm} 
m{1.7cm} 
m{3.5cm}} 
\caption{{Summary of research conducted in the \ac{TL}-based application of kidney disease.}}
\label{table:2}\\

\hline
{Ref.}  & {Used method} & {Dataset} & {Methods description} & {Best performance } & {Limitation} \\
\hline
\endfirsthead

\hline
{Ref.}  & {Used method} & {Dataset} & {Methods description} & {Best performance} & {Limitation} \\
\hline
\endhead

\hline
\multicolumn{6}{r}{\scriptsize\textit{Continued on next page}} \\
\endfoot

\hline
\endlastfoot

{\citep{gaikar2022transfer}} & 
{U-Net+{TL}} & {Private data} &
{Proposed a fully automated 2D attention U-Net model for kidney segmentation on T1 weighted-nephrographic phase contrast enhanced} & {{DC}:89.34$\pm$5.31} & {The model's performance may still be limited by the relatively small size of the dataset} \\
\hline

{\citep{song2022ct2us}} &  {{CNN}+CM +{TL}} & {Private data}  & {Propose a cross-modal \ac{TL} from \ac{CT} to \ac{US} to segment kidney organ based on CycleGAN, U-Net, U-Res, PSPNet, and DeepLab v3+ and \ac{CNN} models.} & {{DC}: 0.853} & {Limited training data for ultrasound Images} \\

\hline

{\citep{davis2021deep}} &  {CNN+TL} & {Private data } & {A 9-layer \ac{CNN} based on the common U-Net. The architecture was developed and tested for the segmentation of non-sclerotic and sclerotic glomeruli of kidney.} & {{F1}: 87.00 \newline {SEN}: 93.00} & {Complexity of model.} \\

\hline

{\citep{li2021endoscopic}} & 
{ VGG16+{TL}} &  {Private data} & {The U-Net was used to extract the kidney stone area. The VGG16 is used as the encoder of U-Net model  to extract the semantic information.} & {{F1}: 97.03} & {Pretrained features may limit model's adaptation to endoscopic image characteristics.} \\

\hline

{\citep{yin2019fully}} & {{TL}+{CNN} +{BDRN}} & {CHOP}  & {Used a boundary distance regression network to learn kidney boundary} & {ACC: 98.90 $\pm$0.6 \newline {DC}: 0.94$\pm$0.03} & {Model struggles with \ac{US} artifacts like noise, shadowing, and distortions.} \\

\hline

{\citep{zheng2019computer}} & { {TL}+{SVM}} & { Private data} & { \ac{SVM} classifiers and \ac{TL} are used to classify diseased and normal kidneys.} & {ACC: 87.00 \newline {SEN}:86.00} & {Small dataset size affecting generalizability and robustness.} \\
\hline

{\citep{ayyar2018harnessing}}  & { DNN+{TL}} & { GCDB} & { Test DNN+\ac{TL} model for classifying normal and abnormal categories of glomeruli} & { {ACC}: 88.23 \newline {SEN}: 88.23} & {Small dataset, \ac{TL} models underperform, limited generalizability.}  \\
\hline

{\citep{zheng2018transfer}}  & { {TL}+{SVM}} & { CHOP}  & { Propose a \ac{TL}-based method and \ac{SVM} model to extract features from \ac{US} kidney images to improve the CAKUT diagnosis in children.} & { ACC:0.87$\pm$2.1 \newline AUC:0.92$\pm$0.7} & {\ac{TL} performance impacted by dataset imbalance and feature complexity.} \\\hline

{\citep{islam2022vision}} & {VGG16} & {Private data} & {Development an intelligent system based on EANet, CCT, Swin transformers and Resnet, VGG16, and Inceptionv3 models to auto-diagnose kidney stones, cysts, and tumors.} & {ACC: 99.30 } & {Imbalanced dataset and limited modalities} \\ \hline

{\citep{ahmed2024identification}} & {VGG16 with \ac{XAI}} & {KUB X-ray images} & {Developed a system for classifying kidney stones using \ac{TL} and \ac{XAI} to enhance transparency and decision-making.} & {ACC: 97.41} & {Lacks fairness and explainability without \ac{XAI} integration.} \\ \hline

{\citep{puttur2024advanced}} & 
{VGG16} & 
{KiTS19 Dataset} & 
{Predict \ac{KC} from \ac{CT} scans, integrating VGG16 with \ac{TL}}. & 
{ACC:97.5} & 
{Limited by dataset variability} \\ \hline

{\citep{kumar2024enhancing}} & 
{VGG16} & 
{Kaggle kidney disease} & 
{\ac{TL} for kidney disease classification.} & 
{ACC:98.00} & 
{Requires large labeled data.} \\ \hline

{\citep{prasher2024vgg16}} & 
{VGG16} & 
{Coronal CT images of kidney stones} & 
{Automated kidney stone diagnosis using \ac{TL} model}. & 
{ACC: 99.70} & 
{Performance depends on the quality and consistency of CT images.} \\ \hline

{\citep{wu2024automated}} & 
{VGG18} & 
{Kidney ultrasound images} & 
{Automated renal tumor classification using deep \ac{TL} on ultrasound}. & 
{SEN: 79.00 \newline SPE:86.00} & 
{Limited by ultrasound image quality and operator skill dependency} \\ \hline

{\citep{goker2024transfer}} & 
{AlexNet, Xception, VGG19, ResNet101} & 
{CT images of kidney stones} & 
{A \ac{TL}-based model for kidney stone classification using CLAHE for enhancement, Laplacian filter for sharpening.}. & 
{ACC: 98.15 \newline Recall: 0.991 } & 
{Limited to CT images, and may not generalize to other image modalities or conditions.} \\ \hline

{\citep{canbay2024privacy}} & 
{Xception, ResNet50, InceptionResNetV2, VGG19} & 
{Priviat data} & 
{Proposed a privacy-preserving \ac{TL} framework for kidney disease detection using \ac{CNN}s and differential privacy}. & 
{ACC: 99.83 \newline Recall 0.9983} & 
{Privacy-aware classification results degrade as noise increases} \\ \hline

{\citep{miao2024development}} & 
{ResNet-18 } & 
{Ultrasound images} & 
{CNN with \ac{TL} segments ultrasound images of histotripsy.}. & 
{DSC:85.} & 
{Performance drops in early stages } \\ \hline

{\citep{kwon2024transfer}} & 
{U-Net} & 
{FIESTA and SSFSE MRI datasets} & 
{\ac{TL} applied for improved kidney segmentation using multi-sequence MRI data.}. & 
{DC: 0.952 } & 
{Limited to specific MRI sequences} \\ \hline

{\citep{matos2023evaluation}} & 
{U-Net with EfficientNet as encoder} & 
{KiTS19 Dataset} & 
{Kidney segmentation using \ac{TL} with U-Net architecture for CT images}. & 
{DC:96.00 \newline Jaccard index:94.40} & 
{Limited to kidney tumor segmentation} \\ \hline

{\citep{gido2025kidney}} & 
{nnU-Net} & 
{KiTS19 and UTH datasets} & 
{Explores kidney and tumor segmentation using mixed datasets and fine-tuning \ac{TL}}. & 
{DICE scores:0.964} & 
{Limited to kidney and renal tumor segmentation} \\ \hline

{\citep{krishnan2024deep}} & 
{Modified UNet3 with residual layers} & 
{MRI images of ADPKD kidneys} & 
{\ac{TL} applied to transfer a model trained on human ADPKD \ac{MRI} images to rat and mouse ADPKD models.}. & 
{DSC: 0.93$\pm$0.04} & 
{Class imbalance in target datasets} \\ \hline

{\citep{han2021explainable}} & {Deep convolutional \ac{TL} with SHAP} & {Hunan Cancer Hospital and COIL-100 datasets} & {Proposed an RCC prediction model incorporating SHAP for explainability and DCA for evaluating clinical utility.} & {ACC:73.87 on RCC, \newline ACC:99.81 on COIL-100} & {Lower performance in real-world clinical datasets compared to public datasets.} \\ \hline

{\citep{liao2021data}} & {ACWGAN-GP, MobileNetV2} & {Renal\ac{US} images} & {Developed an auxiliary diagnosis system for KC stage classification using \ac{DL}}. & {ACC: up to 90.1} & {Limited to four stages of KC.} \\ \hline

{\citep{ozbay2024kidney}} & 
{SSLSD-KTD with masked autoencoder} & 
{KAUH-kidney, CT-kidney} & 
{Combines self-distillation, \ac{SSL}, and \ac{TL} for better metrics with small datasets.}. & 
{ACC:99.82} & 
{Limited by dataset size.} \\ \hline

{\citep{rana2024kidneymultinet}} & 
{\ac{CNN}} & 
{Kidney CT scan} & 
{KidneyMultiNet combines DenseNet201 and Xception with \ac{TL} for kidney disease detection.} & 
{ACC: 99.92} & 
{Limited by the dataset} \\ \hline

{\citep{iqbal2025novel}} & 
{\ac{RDAN}} & 
{Real-world KC datasets} & 
{RDAN uses \ac{TL}, MMA, and DML for accurate KC diagnosis.} & 
{ACC: 96.94 \newline Precision: 98.81.} & 
{Requires diverse datasets and may face challenges in clinical deployment across all environments.} \\ \hline

{\citep{liu2024renal}} & 
{Contrastive learning(ResNet-50)} &  
{TG-GATEs pathological image} & 
{Proposed a \ac{TL} model based on contrastive pretraining with mouse glomerulus images to improve human IgA nephritis image classification.} & 
{ACC: 92.22 \newline SEN: 92.74} & 
{Limited to glomerulus detection and classification} \\ \hline

{\citep{kausar2024machine}} & 
{Random Forest, Decision Tree, K-nearest Neighbor.} & 
{Pediatric Ultrasound Kidney Images } & 
{Introduces a BCOA to select optimal features from high-dimensional ultrasound images for KC detection}. & 
{ACC:99.00 \newline
Precision: 100} & 
{The performance of the model may be highly dependent on the dataset size} \\ \hline

{\citep{nagarajan2024ensemble}} & 
{ResNet-50, VGG16, InceptionV3} & 
{KiTS19 Dataset} & 
{Proposes an ensemble \ac{CNN} model with \ac{TL} for kidney segmentation}. & 
{ACC:98 \newline Precision: 0.91} & 
{Performance depends on image quality} \\ \hline

{\citep{sharma2024transfer}} & 
{Fine-tuned InceptionV3} & 
{Priviat data} & 
{Proposes AI system using \ac{TL}(InceptionV3) for autonomous kidney disease diagnosis.}. & 
{ACC: 96.52 \newline
Precision: 97.76} & 
{The model's performance depends heavily on high-quality} \\ \hline

{\citep{kaur2024advancements}} & 
{EfficientNetB3} & 
{Priviat data} & 
{Proposes EfficientNetB3 \ac{TL} model for automatic kidney disease classification.} & 
{ACC:99.00} & 
{The model's performance is influenced by the quality and diversity of the training dataset.} \\ \hline

{\citep{nandhitha2024impact}} & 
{SqueezeNet} & 
{Kidney radiographs} & 
{Analyzing activation functions' impact on SqueezeNet classification}. & 
{ACC: 51.17 \newline SEN:90.55} & 
{Performance depends on activation function selection} \\ \hline

{\citep{kumar2024comprehensive}} & 
{Xception} & 
{Kidney images dataset} & 
{AI-based system for accurate kidney disease prediction and classification.}. & 
{ACC: 99.89} & 
{Requires large, well-labeled dataset for effective training.} \\ \hline

{\citep{ghosh2025fuzzy}} & 
{STREAMLINERS } & 
{Kidney CT image dataset} & 
{Fuzzy logic-enhanced CT images, ensemble learning, \ac{TL} for tumor detection.}. & {ACC: 99.25} & 
{Sensitive to noisy images and data imbalance.} \\ \hline

{\citep{chaki2024efficient}} & 
{DarkNet19, InceptionV3, ResNet101} & 
{Kaggle kidney CT image } & 
{Ensemble model using inductive transfer for kidney stone detection}. & 
{ACC: 99.8} & 
{Limited dataset and performance degradation with noisy images} \\ \hline

{\citep{lopez2023boosting}} & 
{ResNet50 with Two-step \ac{TL}} & 
{Endoscopic and CCD camera images} & 
{Two-step \ac{TL} for kidney stone classification}. & 
{ACC:0.904$\pm$0.048} & 
{Small dataset and class imbalance} \\ \hline

{\citep{andreini2024enhancing}} & 
{MobileNet, SegNeXt} & 
{Human Glomeruli, Mouse Glomeruli} & 
{Cross-species \ac{TL} for kidney glomeruli segmentation}. & 
{Dice score:\newline90.43} & 
{Data differences between species} \\ \hline

\end{longtable}
\begin{flushleft}
\textbf{Note:} The best performance are in (\%).
\end{flushleft}
}

\subsection{Multi-modal \ac{TL}-based \ac{KC}} 

\begin{figure}[ht!]
\begin{center}
\includegraphics[width=0.7\columnwidth]{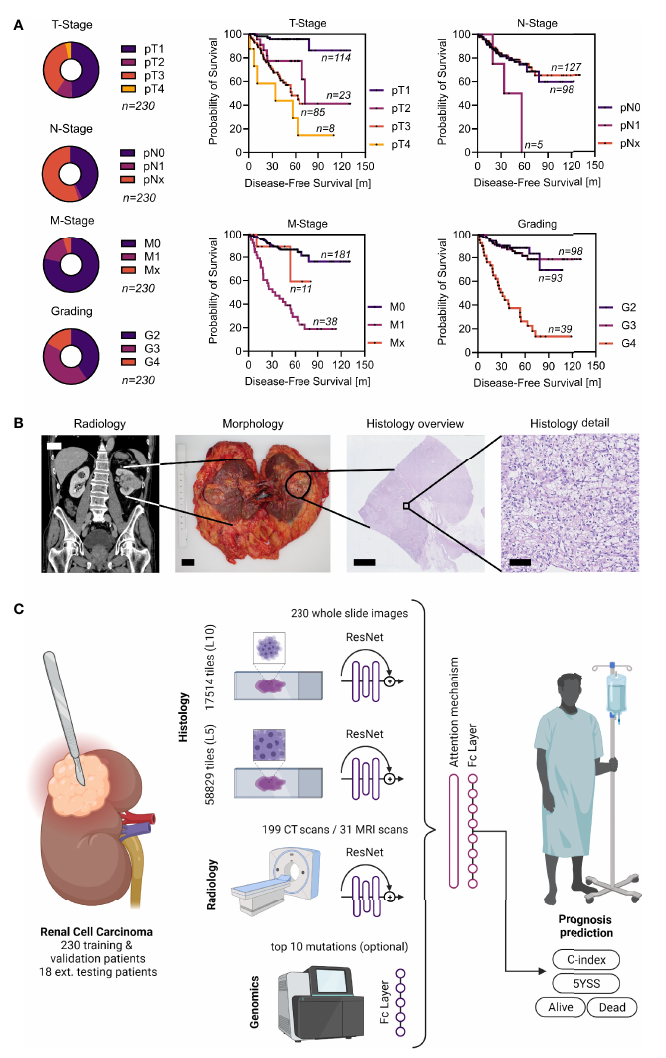}
\end{center}
\caption{An example of the use of multi modal data fusion in cancer prognosis, specifically for \ac{RCC}. This approach integrates diverse data types such as radiological images, histology slides, and genomic information into a unified model to enhance diagnostic and prognostic accuracy. Panel (A) demonstrates survival analysis across different clinical stages (T, N, M stages, and grading), illustrating how integrating various data sources can provide deeper insights into disease progression. Panel (B) shows how radiological, morphological, and histological data are combined to analyze tumor characteristics, while Panel (C) outlines the fusion process, using imaging and genomic data to create predictive models that help in forecasting survival outcomes and guiding treatment strategies \cite{schulz2021multimodal}.}
\label{fig8}
\end{figure}

\ac{TL} has been instrumental in enhancing the precision of \ac{KC} diagnosis and prognosis by consolidating diverse data sources, including \ac{CT} scans, \ac{MRI} images, and gene expression data. This approach allows for a comprehensive analysis by leveraging the strengths of each data type. For instance, imaging data like \ac{CT} and \ac{MRI} scans offer visual insights into the physical state of the kidneys, identifying tumors' size, shape, and location. On the other hand, gene expression data provides a molecular perspective, offering clues about the tumor's aggressiveness and potential response to treatment. By applying \ac{TL}, models pre-trained on large datasets from one domain (e.g., radiology imaging) can be fine-tuned with data from another domain (e.g., genomic data), facilitating a multi-faceted understanding of \ac{KC}. This integration enables the models to learn from the complex interplay between the imaging characteristics of tumors and their molecular profiles, leading to more accurate and personalized diagnostic and prognostic assessments (Figure \ref{fig8}). The ability to draw upon pre-existing knowledge from various domains not only saves significant time and resources but also enhances the models' predictive capabilities, making \ac{TL} a powerful tool in the fight against \ac{KC}. 

\subsection{Explainable AI for \ac{TL}-based \ac{KC}} 
The potential of \ac{XAI} in the diagnosis of \ac{KC} presents a promising frontier in medical imaging and oncology \citep{vasquez2019explainable}. 
Ahmed et al. \citep{ahmed2024identification} address this by applying a \ac{TL} model with a pre-trained VGG16 architecture, enhanced with \ac{XAI} techniques like {\ac{LRP}}, achieving a remarkable accuracy in classifying KUB X-ray images as kidney stones or normal cases. This incorporation of \ac{XAI} not only boosts the model's performance but also its transparency, making the decision-making process more understandable to medical professionals. In the same direction, Han et al. (2021) tackle the prediction of \ac{RCC}  using deep convolutional \ac{TL} and SHAP for model interpretability, achieving significant accuracies and providing valuable insights into model decisions \citep{han2021explainable}.  This approach not only facilitates the understanding of how certain features affect the model’s prediction but also improves the clinical utility of the models by providing insights that are actionable in clinical practice.

\section{Existing datasets and metrics }\label{sec5}
The efficacy of \ac{CAD} systems is significantly dependent on the dataset they are trained on, whether these systems employ \ac{TL} techniques or other methodologies. The volume and integrity of the dataset not only affect its quality but also the performance of the resultant models. Yet, the acquisition of extensive medical imaging data presents challenges due to the intricate nature of imaging techniques and potential risks to patients. Moreover, the process of annotating such data demands the expertise of skilled radiologists to ensure precision, making high-quality, large-scale annotated datasets rare. In response to this limitation, research into novel data augmentation strategies, including photometric and geometric alterations, is underway to enrich existing datasets, thereby alleviating the issue of data scarcity. Enhancing the dataset's quality via advanced imaging technique is another strategy to bolster the performance of models based on \ac{TL}. These initiatives pave the way for enhanced model outcomes through the use of enriched data. There are several key datasets accessible to the public that offer insights into kidney tumors, namely: (i) The \ac{KiTS19} challenge dataset, offering \ac{CT} scans for tumor segmentation algorithm assessment. (ii) The national cancer institute's kidney renal clear cell carcinoma dataset, with gene expression and clinical details pertinent to {\ac{CCRCC}}, the prevalent form of \ac{KC}. (iii) The \ac{TCGA} kidney renal papillary cell carcinoma dataset, which provides genomic and clinical data for patients with {\ac{PRCC}}. (iv) The dataset from radboud university medical center focusing on kidney tumors, featuring \ac{CT} scans alongside segmentation masks for evaluating kidney tumor segmentation techniques. Each dataset is distinct in terms of the data it encompasses, the patient count, and the imaging techniques employed, as summarized in Table.\,\ref{table:5}, which centers on datasets related to kidney tumors.

\begin{scriptsize}
\begin{longtable}{m{0.5cm} m{1cm} 
m{1cm}m{1cm}m{0.5cm}m{0.5cm}m{8.5cm}m{2cm}}
\caption{Summary of the most frequently used KC datasets.} \label{table:5} \\
\hline
\textbf{Ref} & \textbf{Available?} & \textbf{Datasets} & \textbf{Format} & \textbf{RR} & \textbf{Year} & \textbf{Applications} & \textbf{The statistics of dataset} \\
\hline
\endfirsthead

\hline
\textbf{Ref} & \textbf{Available?} & \textbf{Datasets} & \textbf{Format} & \textbf{RR} & \textbf{Year} & \textbf{Applications} & \textbf{The statistics of dataset} \\
\hline
\endhead

\hline
\endfoot

\hline
\endlastfoot

\citep{link1} & Yes\footnote{\url{https://archive.ics.uci.edu/ml/datasets/chronic_kidney_disease}} & UCIMLR & DICOM & \citep{amirgaliyev2018analysis} & 2018 & Classify KC using SVM & 400 cases \newline 25 attributes \\

& & & & \citep{almansour2019neural} & 2019 & Diagnosis of KC using ANN and SVM & 400 patients \newline 24 attributes \\

& & & & \citep{jerlin2020efficient} & 2020 & Classify KC using MKSVM and FFOA & 400 instances \newline 25 attributes \\

& & & & \citep{hsiao2021automatic} & 2021 & Predict KC using TL & 300 patients \newline 210 labels \\

& & & & \citep{goyal2022automated} & 2022 & Kidney segmentation using RCNN & -- \\

\citep{link2} & Yes\footnote{\url{https://www.cancer.gov/tcga}} & TCGA & DICOM & \citep{kocak2018textural} & 2018 & Distinction between RCC using ANN and SVM & 26 patients \\

& & & & \citep{kocak2019radiogenomics} & 2019 & Prediction of mutation in RCC using ANN and RF & 267 patients \\

& & & & \citep{lin2020ct} & 2020 & Differentiating high- from low-grade RCC using TL & 390 patients \\

& & & & \citep{huang2021exploration} & 2021 & Integrate CT and genomics features for RCC using SVM-RFE & 205 patients \newline 107 features \\

& & & & \citep{wu2022comprehensive} & 2022 & Classify genomic characteristics in RCC using ML & 575 cases \\

\citep{link3} & Yes\footnote{\url{https://kits19.grand-challenge.org/}} & KiTS19 & DICOM \newline NIfTI & \citep{graham2019accurate} & 2019 & Segmentation of kidneys from CT images of kidney stone patients & 153 patients \\

& & & & \citep{da2020kidney} & 2020 & Delimit the kidneys in CT images & 210 patients \\

& & & & \citep{hsiao2021automatic} & 2021 & Kidney segmentation & 300 patients \newline 210 labels \\

& & & & \citep{zhu2022improving} & 2022 & Segment kidney and tumor regions & 20 patients \newline 210 images \\

\citep{link4} & Yes\footnote{\url{https://www.ircad.fr/research/data-sets/}} & IRCAD & DICOM \newline NIfTI & \citep{xia2019deep} & 2019 & Renal segmentation & 363+128 patients \\

\citep{link5} & Yes\footnote{\url{http://www.miccai.org/}} & MICCAI & DICOM & \citep{efremova2019automatic} & 2019 & Kidney segmentation & 210 patients \\

\end{longtable}

\begin{flushleft}
Abbreviation: Relevant references (RR)
\end{flushleft}
\end{scriptsize}

When discussing the advantages and limitations of the DL and TL frameworks, it is essential to specify the criteria used for their evaluation, including key performance metrics such as accuracy, precision, recall, F1 score, and computational efficiency (e.g., training and inference time, processing power requirements). These metrics provide a comprehensive assessment of the models' predictive capabilities, balancing correctness (accuracy), sensitivity (recall), and the trade-off between precision and recall (F1 score). In the context of \ac{TL}, selecting the right evaluation metrics is particularly crucial for understanding how well a pre-trained model adapts to a new domain and generalizes after transfer. This comparative analysis can highlight where the proposed frameworks excel and where improvements may be needed. Additionally, conducting comparative quantitative performance evaluations would be valuable, allowing direct comparisons with state-of-the-art methods and providing a clearer understanding of the frameworks' relative strengths and weaknesses. Table \ref{table:6} provides detailed explanations of commonly used evaluation metrics in \ac{TL}, further aiding the assessment of their effectiveness in real-world applications.

\begin{small}
\begin{longtable}{m{4.5cm}m{3cm}m{8cm}}
\caption{Detailed summary of evaluation metrics for \ac{TL}} \label{table:6} \\

\hline
\textbf{Metric} & \textbf{Formula} & \textbf{Detailed Insight} \\
\hline
\endfirsthead

\hline
\textbf{Metric} & \textbf{Formula} & \textbf{Detailed Insight} \\
\hline
\endhead

\hline
\endfoot

\hline
\endlastfoot

\textbf{Accuracy} & $\frac{\text{TP} + \text{TN}}{\text{TP} + \text{FP} + \text{TN} + \text{FN}}$ & Measures the proportion of correct predictions (both positives and negatives). Simple but not always reliable for imbalanced data. \\
\hline
\textbf{Recall (True Positive Rate)} & $\frac{\text{TP}}{\text{TP} + \text{FN}}$ & Measures how well the model identifies positive instances. Important when missing a positive instance (false negative) is critical. \\
\hline
\textbf{Precision} & $\frac{\text{TP}}{\text{TP} + \text{FP}}$ & Measures the accuracy of the positive predictions. Important when false positives (incorrect positives) are costly. \\
\hline
\textbf{F1 Score} & $2 \times \frac{\text{Precision} \times \text{Recall}}{\text{Precision} + \text{Recall}}$ & Combines precision and recall into a single metric. Useful when there's a need to balance both, especially with imbalanced datasets. \\
\hline
\textbf{Specificity (True Negative Rate)} & $\frac{\text{TN}}{\text{TN} + \text{FP}}$ & Measures the proportion of correctly identified negative instances. Important in scenarios where false positives are undesirable. \\
\hline
\textbf{AUC - ROC} & \text{Area under ROC curve} & Measures the model's ability to distinguish between positive and negative classes. A higher AUC value indicates better discriminative ability across various thresholds. \\
\end{longtable}
\end{small}

\section{Limitations and key challenges}\label{sec7}

\ac{TL} can have several advantages in the context of \ac{KC}, including 
securing annotated data in the healthcare sector is a significant challenge, often requiring considerable effort. \ac{TL} plays a pivotal role in this context by facilitating the use of existing pre-trained models. These models, which have been educated on large and universally accessible datasets, can be adapted to develop specialized models for diagnosing specific conditions like \ac{KC}. The advantages of applying pre-trained models are manifold. Firstly, these models, having been exposed to vast amounts of data, have the capability to significantly enhance performance on smaller, task-specific datasets, particularly in diagnosing \ac{KC}. This approach not only leads to considerable savings in time and financial resources but also accelerates the training process. This is because fine-tuning a pre-trained model for a new task is much faster than starting training from scratch. Furthermore, \ac{TL} reduces the need for extensive data and computational power for training models on specific tasks. In the realm of medical imaging, where missing or scarce data can pose significant challenges, especially for rare diseases like \ac{KC}, \ac{TL} offers a valuable solution by employing models that have been pre-trained on extensive datasets. This method effectively bridges the data gap, providing a way to address the issues of scarce data and domain shift that are prevalent due to the high variability in medical imaging data, making it especially beneficial for scenarios involving small datasets.

\noindent In addition there are several limitations to using \ac{TL} for \ac{KC} diagnosis:

The application of \ac{TL} in diagnosing \ac{KC} presents several challenges and considerations. One primary challenge is {\ac{DA}}, which necessitates identifying a pre-trained model trained on a domain or task closely related to \ac{KC} diagnosis. The effectiveness of \ac{TL} hinges on the alignment between the original and target domains and tasks, with discrepancies potentially undermining the model's ability to transfer knowledge effectively. Moreover, the risk of overfitting is heightened when fine-tuning pre-trained models on limited datasets, leading to models that perform poorly on new or unseen data. Fine-tuning entails careful adjustments of model layers, learning rates, and regularization methods to avoid such overfitting. Data scarcity for the new task can further complicate fine-tuning, potentially resulting in suboptimal performance. Additionally, \ac{TL}  can be computationally demanding, especially with large pre-trained models, posing scalability challenges. Interpretability issues also arise, as understanding the decision-making process of a model trained on a different task can be complex. The biases in the datasets used for training pre-trained models may carry over, affecting performance on certain data subsets. Moreover, these models, being trained on specific datasets, might not generalize well across different datasets, especially when data distributions vary significantly. Privacy concerns are another critical consideration, particularly when sensitive medical data must be shared with third parties for model training. The architecture of the pre-trained model also plays a crucial role in the success of \ac{TL} ; an ill-suited architecture can hinder effective fine-tuning. Lastly, the substantial computational resources required for \ac{TL}  can make the use of pre-trained models costly, especially when employing multiple models or conducting several rounds of fine-tuning. These challenges underscore the complexity and nuanced considerations involved in applying \ac{TL}  for medical diagnosis tasks such as \ac{KC}.

The studies reviewed in this article emphasize the effectiveness of \ac{TL} for diagnosing \ac{KC}, noting its computational efficiency and potential to outperform conventional \ac{ML} and \ac{DL} techniques. However, several key challenges still need to be addressed to enhance the performance of \ac{TL}. This section aims to highlight the most significant unresolved issues and current concerns that are receiving considerable focus in the area of \ac{KC} diagnosis:

\subsection{Availability, adapting and adjusting models} 
Data accessibility plays a pivotal role in crafting and applying \ac{TL} models for diagnosing \ac{KC}. Nevertheless, several challenges complicate data accessibility in this arena. A primary issue is the scarcity of annotated medical imaging data specific to \ac{KC}. This data, essential for \ac{TL} model training, comprises images marked with details indicating the presence or absence of diseases like \ac{KC}, yet it's often limited in both amount and quality. Another obstacle arises from the inconsistencies in imaging techniques and data collection methods among various hospitals and healthcare facilities, which can introduce irregularities in the data for \ac{TL} models, adversely affecting their performance and precision. Moreover, concerns related to privacy and security pose significant restrictions on medical imaging data's availability for \ac{TL}, given the sensitive nature of patient information embedded in such data, complicating its distribution among distinct entities. Additionally, data bias presents a critical challenge; \ac{TL} models trained on skewed data might exacerbate existing inequalities in medical diagnoses and treatments. For instance, a model trained predominantly on data from male subjects might underperform with female patient images, reflecting and potentially worsening disparities.

The primary obstacle in \ac{TL} involves identifying an appropriately pre-trained model that has been trained on a domain or task closely related to the new task at hand. It's critical to evaluate how similar the original (source) and the intended (target) domains are, as well as the resemblance between the tasks involved \citep{guan2021domain, gadermayr2019generative}.

Adjusting a pre-trained model for a specific task, known as fine-tuning, presents its own set of challenges. It necessitates a strategic selection of which layers to adjust, alongside the determination of an optimal learning rate and the application of appropriate regularization strategies to prevent the model from becoming too narrowly adapted to the task \citep{uldry2017fine}.

\subsection{Data scarcity}
Data scarcity presents a significant challenge in the development and efficacy of {\ac{KC}} diagnosis systems, impacting everything from image analysis to the training of {\ac{DL}} models. Several studies have explored these challenges and proposed methods to mitigate the impact of limited data availability \citep{upadhyay2023advances}.
Peng et al. \citep{peng2019extent} explore how downsampling and compression affect renal image analysis. They found that data scarcity impacts diagnostic accuracy, highlighting the need for robust data handling techniques to ensure the integrity and usefulness of renal images in diagnosis. Similarly, Guo et al. \citep{guo2022deep} address the limited training data for kidney segmentation by using a cascaded {\ac{CNN}}, demonstrating that innovative network architectures can partly overcome the challenges posed by sparse data.

In the context of clinical applications, Heller et al. \citep{heller2019kits19} provide the kits19 dataset, which includes 300 kidney tumor cases with CT semantic segmentations and surgical outcomes, aiming to enhance research in {\ac{KC}} by making rich datasets available. This effort underscores the critical role of accessible high-quality datasets in advancing early tumor detection and improving surgical outcomes.
Flitcroft et al. \citep{flitcroft2022early} and Nassour et al. \citep{nassour2023relative} both discuss the scarcity of reliable diagnostic biomarkers for {\ac{KC}}, particularly in non-invasive strategies like urinary protein analysis and the evaluation of Lynch syndrome’s impact on {\ac{KC}}. These studies highlight the gap in epidemiological data and the need for comprehensive datasets to develop effective diagnostic tools.

Technological advancements in handling data scarcity are also evident in {\ac{DL}} applications. Alzu'bi et al. \citep{alzu2022kidney} and Bhattacharjee et al. \citep{bhattacharjee2023multi} discuss the use of CNNs for kidney tumor classification and detection. They focus on optimizing {\ac{DL}} approaches to perform effectively even with limited data, a crucial advancement given the rarity and variability of KC data.
Lastly, Wang et al. \citep{wang2023deep} delve into the utilization of {\ac{DL}} techniques for the imaging diagnosis of \ac{RCC}. They emphasize how these advanced computational tools allow clinicians to diagnose and evaluate renal tumors more accurately and swiftly. One of the key strategies discussed is leveraging unlabeled data, which is a valuable solution for coping with data scarcity. By utilizing semi-supervised learning techniques, the potential of existing datasets can be maximized without the need for extensive labeled data.

\subsection{Data bias and generalization} 
Data bias represents a significant obstacle in \ac{TL} for \ac{KC} diagnosis, impacting model accuracy and fairness. Bias can manifest in various forms, such as sampling bias, where training data may not accurately reflect the disease's true distribution, potentially leading to population-specific biases \citep{zhou2019quality}. Measurement bias skews data collection towards specific attributes, like age or gender, causing model predictions to favor these characteristics. Annotation bias arises from annotator prejudices, affecting the reliability of data labels \citep{harrison2021risk}. To counteract these issues, \ac{TL} models may utilize data augmentation to enrich training data variety and volume. Training models on data from diverse sources, like multiple hospitals, ensures a representative dataset of the disease's distribution. Additionally, implementing fairness constraints and promoting algorithmic transparency can help mitigate bias, fostering the development of fair and unbiased models \citep{huang2020chronic}.

Generalization stands as a pivotal challenge in \ac{TL} for diagnosing \ac{KC}, denoting the model's capacity to deliver accurate predictions on unseen data. Achieving robust generalization is essential, determining the model's viability in practical scenarios \citep{pirmoradi2021self}. A notable hindrance to strong generalization is the limited availability of annotated data. For many conditions, including \ac{KC}, annotated data for training may be scarce, leading to a risk of overfitting, where the model overly specializes on training data, diminishing its performance on novel data. Additionally, data diversity issues, where training data may not accurately represent the disease's distribution, can impair generalization \citep{van2022diagnosis}. For instance, data sourced from a singular hospital might not mirror the broader disease prevalence across different locales. To mitigate these issues, \ac{TL} models for \ac{KC} diagnosis may incorporate data augmentation strategies to broaden the training data's size and variety. Training on data from varied sources, like multiple healthcare facilities, ensures a more representative dataset. Furthermore, employing regularization methods like dropout aids in curbing overfitting, enhancing the model's generalization capabilities \citep{shevchenko2020evaluation}.

\subsection{Privacy concerns, expertise shortage and  accuracy instability} 
Privacy concerns emerge prominently in developing and applying \ac{TL} models for \ac{KC} diagnosis due to the necessity of accessing sensitive medical information, triggering significant privacy and security considerations \citep{zhou2019quality}. A key challenge is safeguarding medical data privacy and security, demanding stringent measures against unauthorized access and potential breaches. Managing medical data with utmost care to align with privacy regulations is crucial \citep{nasir2022kidney}. The annotation process also presents privacy challenges, necessitating judicious annotator selection and bias minimization efforts. Employing privacy-preserving methods, such as differential privacy, is essential to protect patient confidentiality during annotation. Additionally, securing \ac{TL} models against unauthorized interventions and ensuring their resilience is paramount. Providing patients with transparent data usage information and granting them control over their data usage decisions is vital for maintaining trust and privacy integrity \citep{harrison2015patients}.

The creation and implementation of \ac{TL} models for \ac{KC} diagnosis necessitate profound expertise in medical imaging and \ac{ML}, yet expertise scarcity poses several challenges \citep{omotoso2023addressing}. The limited pool of experts hampers the resources and knowledge necessary for these models' development and application. Interdisciplinary collaboration, essential between medical and \ac{ML} experts, is often obstructed by differing methodologies and communication barriers \citep{ferrari2019defining}. Adhering to medical imaging and m\ac{ML} best practices and standards is imperative for the responsible development and deployment of \ac{TL} models. Ensuring patient safety and efficacy of these models requires a deep understanding of patient needs and the development of models that positively impact patient outcomes.

Improving \ac{TL} model performance often involves utilizing pre-trained models on extensive datasets, such as those trained on the ImageNet database. Nonetheless, limited data availability can lead to accuracy instability. Various regularization techniques, such as SPAR, have been explored to stabilize accuracy by moderating the target network's outer layer weights with reference to a starting point \citep{arnold2016duration}. An innovative regularization method, \ac{DL} transfer with attention, was introduced, focusing on preserving the source model's outer layer outputs rather than merely constraining the model's weights. \ac{DL} transfer with attention minimizes empirical loss and aligns the two models' outer layer outputs by selectively constraining feature maps identified by a supervised learning-trained attention mechanism .

\subsection{Knowledge gain and data annotation}
Evaluating the knowledge gained by the \ac{TL} model and its performance outcomes is crucial. Consequently, considerable research has been directed towards establishing precise metrics for assessing the effectiveness of these models and the extent of knowledge they acquire, aiding in the elucidation of complex issues \citep{huang2020chronic}. Proposed metrics include transmission error, transmission loss, transmission ratio, and in-range ratio. Nevertheless, the specificity of these metrics often falls short of expectations, highlighted by the ongoing uncertainty regarding outcomes when other \ac{TL}-based models are employed, such as in \ac{KC} detection scenarios, where model performance might be suboptimal.

Data annotation plays an essential role in developing \ac{TL} models for diagnosing \ac{KC}. Yet, this process faces several significant hurdles. A primary issue is the considerable cost and time required to annotate extensive datasets accurately. The process demands substantial expert input and resources, making it both time-intensive and costly. Another obstacle is the absence of standardized practices in data annotation, complicating the comparison and amalgamation of datasets from diverse origins and potentially diminishing the \ac{TL} models' impact. Ensuring the annotations' quality and precision further necessitates meticulous validation and quality control, adding to the resource and time expenditure. Moreover, achieving representativeness in annotations, ensuring they accurately reflect the target demographic, involves careful annotator selection and concerted efforts to minimize bias throughout the annotation process.

\subsection{Negative transfer and overfitting}
Negative transfer in the context of learning about \ac{KC} refers to the difficulties that individuals may experience when acquiring new knowledge or skills related to the condition and its treatment \citep{ayana2021transfer, chui2023multiround}. (1) Overgeneralization: Individuals may apply learned information too broadly, which may result in an incorrect or incomplete understanding of the condition and its treatment. (2) Overreliance on prior knowledge: Individuals may rely too heavily on prior knowledge about cancer or other medical conditions, which may not be applicable to \ac{KC}.(3) Incompatibility between old and new learning: Prior learning about cancer or other medical conditions may be incompatible with new information about \ac{KC}, leading to confusion and hindering the acquisition of new knowledge. (4) Resistance to change: Individuals may be resistant to changing their existing beliefs and attitudes about cancer, leading to difficulty in adapting to new information about \ac{KC}. (5) Misapplication of prior knowledge: Individuals may misapply prior knowledge, leading to an incorrect or inefficient understanding of the condition and its treatment.

Overfitting, especially in the domain of \ac{KC} diagnosis, describes the situation where a predictive model becomes excessively tailored to the training dataset, undermining its capability to perform accurately on new, unseen data. This often results from training a model with an overly complex set of parameters, rendering it over-specialized and ineffective for general prediction or diagnosis purposes \citep{mutasa2020understanding, azuaje2019connecting}. In the realm of \ac{KC}, overfitting may result in inaccurate diagnoses, an increase in false positives, or diminished disease detection sensitivity, posing significant risks to patient management and therapy. Implementing measures such as regularization to curb model complexity and employing cross-validation to evaluate model efficacy on novel data are critical strategies to prevent overfitting.

The study by \citep{li2020analyzing} introduces innovative regularization techniques, including large margin loss, focal loss, adversarial training, mixup, and data augmentation. These methods aim to correct the logit shift in underrepresented classes and address class imbalance by scrutinizing network behavior, providing a robust approach to mitigate overfitting.

Furthermore, \citep{nikpanah2021deep} details the use of Five-fold cross-validation as a means to assess the performance of \ac{AI} algorithms, specifically in the segmentation process involving seed placement and bounding box techniques for lesion patch extraction, coupled with deep \ac{CNN} for distinguishing clear \ac{CCRCC} from renal oncocytoma.

\section{Future research directions}\label{sec8}
Despite achieving high accuracy, prior studies highlight the need for improved determination of \ac{KC}. Researchers suggest employing less conventional algorithms to enhance early-stage detection. Enhancing \ac{TL}-based models for robust, unbiased \ac{KC} identification across stages is crucial. Integrating feature selection with \ac{DL} models and expanding datasets from various sources may support this goal. Further refinement of algorithms for related classification tasks is also needed. Potential future directions for \ac{TL} in \ac{KC} diagnosis are outlined in the following subsections.

\subsection{Blockchain technology}
In the context of \ac{KC} diagnosis, \ac{TL} plays a vital role in enhancing the effectiveness of blockchain technology. \ac{TL} allows \ac{DL} models to be trained on large, diverse datasets of medical images, such as \ac{CT} scans or \ac{MRI}s, for detecting tumors. Initially, the model is trained on a broad range of cancer data, and later refined using a more specialized dataset focused on \ac{KC}. This process helps the model to become highly accurate in detecting \ac{KC}-specific characteristics. Blockchain technology complements \ac{TL} by providing a secure, transparent, and immutable platform to store and share sensitive medical data. It ensures that patient records are not only protected but also verifiable, offering patients control over their data and granting permissions to authorized medical professionals for analysis. Through blockchain’s decentralized nature, researchers and healthcare institutions can securely share and collaborate on medical data, enabling the use of \ac{TL}-based models across multiple centers. This integration enhances diagnostic accuracy, reduces bias, and improves patient care outcomes by allowing for more precise, timely, and collaborative diagnoses of kidney cancer \citep{himeur2022blockchain, nasir2022kidney}.

\subsection{Adversarial training} 
The burgeoning field of adversarial training offers a compelling avenue for enhancing the diagnosis, prediction, and segmentation of \ac{KC}. This approach leverages the strengths of generative adversarial networks (GANs) and adversarial training methods to address some of the inherent challenges in medical imaging, such as data scarcity, variability in tumor appearance, and the robustness of diagnostic models against adversarial examples. Furthermore, \ac{TL} plays a pivotal role in overcoming the challenge of limited labeled data in medical imaging by enabling the use of pre-trained models on large-scale datasets, which can be fine-tuned for specific tasks like \ac{KC} detection. \ac{TL} allows models to capture essential features from general medical imaging datasets and adapt them to the nuances of \ac{KC}, thus improving diagnostic performance and model generalization. This integration of adversarial training with \ac{TL} has the potential to significantly enhance the accuracy and robustness of \ac{KC} diagnosis models, especially in clinical environments with limited annotated datasets \cite{noureddine2023adversarial, shan2023automatic, zeng2021accurate, brinda2023chronic}.

\subsection{Compression and denoising based on \ac{TL} for \ac{KC}}
A future direction for improving \ac{KC} diagnosis and treatment could involve further advancements in \ac{TL}. These innovations could enhance both medical image compression and tumor detection processes. Specifically, refining \ac{TL} techniques for compressing \ac{KC} images could ensure that critical diagnostic information is preserved while reducing computational and storage demands. Future research could explore more advanced model compression strategies, dynamic loss functions, and real-time data sharing to support collaborative decision-making among healthcare professionals. By combining these strategies, the goal would be to create faster, more accurate, and scalable solutions for detecting and diagnosing \ac{KC}, leading to better patient care and treatment outcomes \citep{habchi2023new, beladgham2019medical, habchi2022improving, fangxing2023improved}. Moreover, employing efficient pretrained denoising \ac{DL} techniques—such as the one presented in \citep{boucherit2025reinforced}—can significantly improve the quality of medical images prior to analysis, thereby enhancing the accuracy and reliability of \ac{KC} detection and diagnosis.

\subsection{Quantum computing}
Quantum computing represents a nascent and burgeoning domain poised to transform our approach to computational challenges, including those within \ac{ML}, with the potential to revolutionize various aspects of knowledge discovery. Despite being at an early developmental stage, quantum computing harbors the promise of significantly enhancing the efficacy of \ac{TL} models in the diagnosis of diseases like \ac{KC} \citep{balamurugan2024optical}. Employing quantum computing in \ac{TL} for \ac{KC} diagnosis entails harnessing quantum algorithms to tackle tasks that would otherwise be impractical with classical computers, notably optimization and linear algebra computations. These algorithms offer avenues to bolster \ac{TL} model performance by curtailing computational overheads and enhancing model accuracy. Additionally, quantum computing holds promise in mitigating classical computing constraints encountered in medical imaging, such as the demand for substantial computational resources and memory for processing voluminous datasets. By leveraging quantum computing, the duration required for training and inference can be slashed, fostering swifter diagnoses and improved patient outcomes \citep{prajapati2023quantum,montiel2023quantum, sahib2023investigation}.

\subsection{DA for clinical data}
{\ac{KC} originates in the kidneys and is commonly detected through imaging techniques like \ac{CT} scans, \ac{MRI}, or \ac{US}, often presenting as a mass or lesion. The prognosis largely depends on the stage at diagnosis, with early-stage cancers having a better survival rate. However, \ac{TL} models used for clinical data often face challenges due to differences in imaging modalities, data distributions, and patient demographics across hospitals or regions. These variations can significantly hinder the performance of models when they are transferred from one domain (e.g., one hospital or imaging technique) to another. \Ac{DA} techniques can address these issues by adjusting pre-trained models to better handle the disparities between datasets, thus improving their ability to generalize across diverse clinical settings. This enables the model to perform well on data from different medical centers, despite differences in the quality and characteristics of the data. These techniques have proven to be particularly useful in scenarios where data from a single institution is limited, as they enhance the model’s robustness and accuracy. By improving model performance across diverse patient populations and imaging modalities, \Ac{DA} helps ensure more reliable and precise predictions, leading to better diagnostic accuracy and patient outcomes. This is crucial in the clinical setting, where consistency and accuracy are paramount for patient care and decision-making \cite{chen2024adversarial}.}

\subsection{\acs{UL}, and \ac{SSL}-based  \ac{TL} for \ac{KC}}
{\Ac{UL}, \ac{SSL}, and \ac{TL} for \ac{KC} diagnosis can provide substantial benefits by enhancing diagnostic accuracy with limited labeled data. \ac{UL} can uncover hidden patterns and clusters in medical images, identifying previously unnoticed features that distinguish cancerous tissues. \ac{SSL} can utilize large amounts of unlabeled data by generating pseudo-labels through pretext tasks, reducing the need for manually annotated datasets. \ac{TL} further amplifies these techniques by enabling models trained on one dataset to be fine-tuned and adapted for \ac{KC} detection in different clinical settings or patient populations, ensuring better generalization and scalability. Together, these methods can significantly improve early detection, reduce the reliance on expert annotation, and enable the development of more robust, efficient, and scalable \ac{AI} systems for \ac{KC} diagnosis \cite{ozbay2024kidney, hu2024one}.}

\subsection{Federated \ac{TL} for \ac{KC}} 
A promising future research direction involves advancing the integration of \ac{TL} and \ac{FL} to tackle the complex issue of identifying kidney abnormalities in medical imaging. This approach could enhance diagnostic capabilities within decentralized healthcare systems while maintaining patient data privacy. Future studies may focus on refining the federated \ac{TL} architecture to enable collaborative learning across geographically dispersed renal imaging datasets, allowing healthcare providers to share insights without compromising data ownership or confidentiality. Further exploration of this methodology could lead to improved precision in renal abnormality detection by adapting \ac{TL} models to better capture the unique characteristics of renal imaging data. Additionally, continued research could enhance the efficiency and security of this system, facilitating more effective, privacy-preserving, and decentralized healthcare solutions. By building on this convergence of deep learning, federated learning, and \ac{TL}, future work could significantly transform the landscape of medical image analysis, fostering a new era of collaborative, secure, and adaptive healthcare \citep{himeur2023federated, vekaria2024identification}.

\subsection{Self-paced learning for KC}
\ac{TL} has proven to be a powerful approach for improving medical diagnostic systems, especially in the complex field of \ac{KC} diagnosis. \ac{TL} leverages knowledge gained from pre-trained models on large datasets to enhance the performance of models on specific tasks, such as \ac{KC} detection. By incorporating a strategy that focuses on progressively more complex or informative cases as the model improves, \ac{TL} adapts to the nuances of the data, helping to address challenges like data variability and imbalance. This adaptive learning process can greatly benefit \ac{KC} diagnosis, allowing the model to effectively handle the complexities of medical imaging and improve diagnostic accuracy over time \cite{yang2019mspl}.

\subsection{Tranformers and Large language models for KC}
The integration of transformer architectures, such as \ac{BERT} and \ac{CTC} among others \citep{djeffal2023automatic,djeffal2024transformer}, into the realm of KC diagnosis and segmentation represents a cutting-edge approach that holds significant promise for advancing medical imaging analysis. Drawing insights from several studies, such as those referenced, the potential applications of transformers in this domain are evident.
One of the primary advantages of transformers lies in their ability to capture long-range dependencies within the data, making them particularly suitable for tasks involving complex relationships and patterns, as often encountered in medical imaging. The referenced studies showcase various transformer-based models tailored specifically for KC diagnosis and segmentation.

For instance, the utilization of convolution-and-transformer networks (COTRNet) proposed by Shen et al. \citep{shen2021automated} offers an end-to-end solution for kidney tumor segmentation, demonstrating the effectiveness of combining convolutional and transformer-based architectures in this context. Similarly, Islam et al. \citep{islam2022vision} leverage vision transformers for the automatic detection of kidney cysts, stones, and tumors from CT radiography, highlighting the versatility of transformer models across different types of kidney abnormalities.

Furthermore, the study by Wang et al. \citep{wang2022combining} showcases the synergy between transformers and {\ac{CNN}s} for renal parenchymal tumors diagnosis. By stacking transformers and {\ac{CNN}}s, the model efficiently captures both long-range dependencies and low-level spatial details, enhancing its performance in identifying and diagnosing kidney lesions.
Moreover, transformer-based methods, such as TA-UNet3+ proposed by Hu \citep{hu2023ta}, introduce innovative attention mechanisms and encoder-decoder architectures to improve kidney tumor segmentation accuracy. Similarly, Qian et al. \citep{qian2023hybrid} present a hybrid network combining nnU-Net and Swin Transformer, emphasizing the critical role of precise delineation and localization in {KC} diagnosis and treatment planning.
Additionally, transformers exhibit promise in histopathological image analysis for subtype classification of \ac{RCC}. Gao et al. \citep{gao2021instance} demonstrate the effectiveness of instance-based vision transformers in subtyping papillary RCC, providing valuable insights for histopathological diagnosis and personalized treatment strategies.

\Ac{LLM} and generative chatbots hold significant potential for transforming {KC} diagnosis in several ways \citep{khennouche2024revolutionizing,yang2022large}. By integrating vast medical libraries and research data, \ac{LLM} can provide healthcare professionals with up-to-date information, diagnostic suggestions, and treatment options based on the latest studies and clinical guidelines \citep{jiang2023health,singhal2023large}. Moving on, generative chatbots can interpret patients' descriptions of their symptoms in natural language, aiding in the early detection of {KC} by flagging potential warning signs and suggesting further diagnostic tests \citep{lievin2023can,wang2023pre,sohail2023decoding}. Typically, these models can deliver personalized information to patients about {KC}, including explanations of their condition, possible treatment pathways, and what to expect during each treatment option, improving patient understanding and engagement in their care plan \citep{malik2023chatgpt,cao2023accuracy,farhat2023analyzing}.

Besides, \ac{LLM} can analyze vast amounts of data from EHRs to identify patterns or risk factors associated with {KC}, potentially uncovering new insights into its diagnosis and progression \citep{coskun2023can,ce2024exploring}. With training, \ac{LLM} can assist radiologists by providing a second opinion on imaging studies, such as CT scans and MRIs, identifying areas that may warrant a closer look for signs of {KC \citep{sultan2023using,srivastava2024utility,al2024traditional}. Additionally, chatbots can handle routine inquiries and patient triaging, allowing medical staff to focus on more complex tasks and patient care, thereby increasing the efficiency of clinical workflows \citep{uprety2023chatgpt,peng2024evaluating}.
\ac{LLM} can integrate and analyze information from various sources, including genomic data, radiographic images, and clinical notes, offering a holistic view of a patient's health status and aiding in the accurate diagnosis of KC  \citep{fink2023potential,sohail2023using}.
Moreover, generative chatbots can facilitate remote monitoring of patient's symptoms and treatment responses, providing timely advice and enabling telehealth consultations with specialists \citep{pugliese2023artificial,sohail2023future}.

\section{Conclusion}\label{sec9}
{The integration of \ac{TL} and \ac{DA} techniques marks a pivotal advancement in enhancing the accuracy, efficiency, and robustness of {KC} diagnosis. \ac{TL} plays a critical role in overcoming the limitations of traditional \ac{DL} approaches by enabling models pre-trained on large, diverse datasets to adapt effectively to smaller, specialized datasets, where annotated data is often scarce—a persistent challenge in medical diagnostics. This knowledge transfer not only reduces the computational burden of training complex models from scratch but also significantly improves diagnostic precision, even when fine-tuning with limited kidney-specific data. Complementing this, \ac{DA} techniques address the critical issue of data distribution shifts, ensuring that diagnostic models remain reliable and accurate amidst the dynamic and evolving nature of medical imaging data. Through a systematic and comprehensive analysis of \ac{TL}-based frameworks for {KC} detection, this study provides a rigorous evaluation of key contributions, identifies pressing challenges, and proposes actionable solutions to push the boundaries of the field. It underscores the transformative potential of \ac{TL} in tackling major hurdles, including the computational complexity of \ac{DL}, the scarcity of annotated medical data, and the variability between training and testing datasets, thereby enhancing diagnostic reliability and precision. Despite these advancements, significant challenges persist. Many \ac{DA} models rely on the assumption of homogeneity in feature distributions between source and target domains—an assumption that often fails to hold in real-world medical scenarios, where data heterogeneity is the norm. Recent research has begun to explore heterogeneous \ac{DA} approaches, which relax these assumptions and open new avenues for innovation. Looking ahead, future research should prioritize the optimization of these heterogeneous \ac{DA} methods to improve their applicability in {KC} detection, fostering the development of more robust, generalizable, and scalable diagnostic systems. Furthermore, efforts should focus on integrating these techniques into clinical workflows, ensuring seamless translation from research to practice. By addressing these challenges, the synergy of \ac{TL} and \ac{DA} has the potential to revolutionize precision medicine, driving transformative changes in clinical practice, improving patient outcomes, and advancing the frontiers of oncology research.}
{However, some limitations of this review must be acknowledged. The studies reviewed primarily focus on a narrow subset of methods, often omitting emerging techniques in the broader \ac{TL} and \ac{DA} fields that may offer significant improvements. Additionally, there is a heavy reliance on studies with limited clinical validation, which could impact the generalizability of the findings to real-world medical environments. Some studies also show methodological inconsistencies in how \ac{TL} and \ac{DA} techniques are applied, making direct comparisons challenging. Furthermore, the review has been constrained by the availability of high-quality, peer-reviewed papers, limiting the inclusion of non-published, yet potentially valuable, research. Future reviews should strive to include a broader range of studies, especially those that explore new, diverse, and interdisciplinary approaches to \ac{TL} and \ac{DA} in medical imaging.}

\section*{Declarations}
\begin{itemize}
\item The authors declare no conflicts of interest.
\item Funding: The authors declare that there is no fund for this project
\item Data availability: There is no data available for this project

\item Ethics approval and consent to participate: 
Not Applicable.

\end{itemize}


\bibliography{references}

\end{document}